\documentclass[structabstract]{aa}  
\usepackage{graphicx}
\usepackage{savesym}
\usepackage{amsmath}
\usepackage{amssymb}
\savesymbol{iint}
\usepackage{txfonts}
\restoresymbol{TXF}{iint}
\usepackage{natbib}
\usepackage{lscape}
\usepackage{rotating}

\usepackage{color}
\begin{document}
   \title{The AMBRE Project: Stellar Parameterisation of the ESO:UVES archived spectra}

   \subtitle{}

   \author{C.~C.~Worley
          \inst{1,2}
          \and
          P.~de~Laverny
	  \inst{1}
          \and
	  A.~Recio--Blanco
	  \inst{1}
          \and
	  V.~Hill
	  \inst{1}
          \and
	  A.~Bijaoui
	  \inst{1}
	  %\inst{2}
	  %\fnmsep\thanks{Thanks to ESO,OCA,CNES}
          } 

   \institute{1. Laboratoire Lagrange (UMR7293), Universit\'e de Nice Sophia Antipolis, CNRS, Observatoire de la C\^ote d'Azur, BP 4229,F-06304 Nice cedex 4, France\\
              \email{ccworley@ast.cam.ac.uk} \\
      2. Institute of Astronomy, University of Cambridge, Madingley Road, Cambridge CB3 0HA, United Kingdom}

% \abstract{}{}{}{}{} 
% 5 {} token are mandatory
 
  \abstract
  % context heading (optional)
  % {} leave it empty if necessary  
   {The AMBRE Project is a collaboration between the European Southern Observatory (ESO) and the Observatoire de la C\^{o}te d'Azur (OCA) that has been established in order to carry out the determination of stellar atmospheric parameters for the archived spectra of four ESO spectrographs.}
  % aims heading (mandatory)
   {The analysis of the UVES archived spectra for their stellar parameters has been completed in the third phase of the AMBRE Project. From the complete ESO:UVES archive dataset that was received covering the period 2000 to 2010, 51921 spectra for the six standard setups were analysed. These correspond to approximately 8014 distinct targets (that comprise stellar and non-stellar objects) by radial coordinate search.}
  % methods heading (mandatory)
   {The AMBRE analysis pipeline integrates spectral normalisation, cleaning and radial velocity correction procedures in order that the UVES spectra can be then analysed automatically with the stellar parameterisation algorithm MATISSE to obtain the stellar atmospheric parameters. The synthetic grid against which the MATISSE analysis is carried out is currently constrained to parameters of FGKM stars only.}
  % results heading (mandatory)
   {Stellar atmospheric parameters are reported for 12,403 of the 51,921 UVES archived spectra analysed in AMBRE:UVES. This equates to $\sim$23.9\% of the sample and $\sim$3,708 stars. Effective temperature, surface gravity, metallicity and alpha element to iron ratio abundances are provided for 10,212 spectra ($\sim$19.7\%), while at least effective temperature is provided for the remaining 2,191 spectra. Radial velocities are reported for 36,881 ($\sim$71.0\%) of the analysed archive spectra. 
   %For 32,306 (62.2\%) spectra, while parameters were determined, they were not considered reliable (and thus not reported to ESO) for reasons such as very low S/N, too poor radial velocity determination, spectral features too broad for analysis, and technical issues from the reduction. Similarly the parameters of a further 7,212 spectra (13.9\%) were also not reported to ESO based on quality criteria and error analysis determined within the automated parameterisation process. It is expected that multi-component stellar systems will return high errors in radial velocity and fitting to the synthetic spectra and therefore, by those tests, mainly not have parameters reported to ESO. 
   Typical external errors of $\sigma_{T_{\textrm{eff}}} \sim 110$~dex, $\sigma_{\log g} \sim 0.18$~dex, $\sigma_{\textrm{[M/H]}} \sim 0.13$~dex, and $\sigma_{[\alpha/\textrm{Fe}]} \sim 0.05$~dex with some reported variation between giants and dwarfs and between setups are reported.}
  % conclusions heading (optional), leave it empty if necessary 
   {UVES is used to observe an extensive collection of stellar and non-stellar objects all of which have been included in the archived dataset provided to OCA by ESO. The AMBRE analysis extracts those objects which lie within the FGKM parameter space of the AMBRE slow rotating synthetic spectra grid. Thus by homogeneous blind analysis AMBRE has successfully extracted and parameterised the targeted FGK stars (23.9\% of the analysed sample) from within the ESO:UVES archive.}


   \maketitle
%
%________________________________________________________________

\section{Introduction}
The development of automated stellar parameterisation routines is in full force in this new era of large scale spectroscopic surveys. In light of current surveys such as RAVE \citep{Steinmetz2006} and the Gaia-ESO Survey \citep{Gilmore2012}, and future surveys such as GALAH \citep{Zucker2012}
and the European Space Agency (ESA) Gaia Mission, having available robust and efficient automated routines that produce reliable parameters and chemical abundances is key to extracting all the potential information that these surveys have to offer.

As outlined in \citet{Worley2012} and \citet{deLaverny2013}, the goal of the AMBRE project is to determine stellar parameters for the archived spectra of four of ESO's high resolution spectrographs: FEROS, UVES, HARPS and Flames/GIRAFFE. A wealth of information remains in the archive spectra outside the goals of the original observing programmes which, in a homogeneous analysis, can make a significant contribution to studies of galactic stellar populations and stellar evolution.

At the basis of this project is the automated parameterisation algorithm MATISSE (MATrix Inversion for Spectral SynthEsis) which has been developed at the Observatoire de la C\^{o}te d'Azur (OCA) for use in the parameterisation of large spectroscopic datasets, in particular for use in the Gaia Radial Velocity Spectrometer (RVS) parameterisation pipeline \citet{Recio-Blanco2016}. The algorithm is fully described in \citet{Recio-Blanco2006}.

As in \citet{Worley2012} for which the parameterisation of the FEROS archive spectra was presented, this paper is devoted to the parameterisation of the UVES archive spectra covering the period from 2000 to 2010. The parameterisation of the HARPS spectra is presented in \citet{DePascale2014}

\begin{table*}[ht]
%TABLE confirmed 22/09/2014
\caption{Characterisation of wavelengths of the six standard UVES setups.}\label{tab:setupinfo}
\tabcolsep=0.11cm
%{\small
\begin{center}
\begin{tabular}{lccccp{3.5cm}p{3.5cm}c}
\hline\hline
Setup & $\lambda$ Range (\AA) & CCD Gap (\AA) & No.Spectra\tablefootmark{a} & No.Target\tablefootmark{b} & Tell.Cont.\tablefootmark{c} $>$5\% \& $<$20\% & Tell.Cont.  $>$20\% & Key Features\\
\hline
BLUE346 & 3043-3916    &        & 5267        &  1687     &                                          &                                         & Balmer Lines \\
BLUE390 & 3281-4612    &        & 11262       &  3949     & 4400-4500\AA                             &                                         & Ca~II H\&K \\
BLUE437 & 3731-4999    &        & 8907        &  2699     & 4400-4500\AA, 4650-4750\AA               &                                         & Ca~II H\&K \\
RED564 & 4583-6686     & 5644-5654 & 11590    &  2467     & 5000-5130\AA, 5370-5520\AA, 5660-5750\AA & 5800-6000\AA,6270-6350\AA, 6450-6610\AA & H$_{\alpha}$,H$_{\beta}$\\
RED580 & 4726-6835     & 5804-5817 & 27912    &  3309     & 5000-5130\AA, 5370-5520\AA, 5660-5750\AA & 5800-6000\AA,6270-6350\AA, 6450-6610\AA & H$_{\alpha}$,H$_{\beta}$ \\
RED860 & 6650-10606    & 8544-8646 & 13468    &  1856     & 6610-6867\AA, 7450-7595\AA               & B-Band(H$_2$O:6867\AA +)                & Ca~II IR Trip.\\
       &               &           &          &           & 7750-7850\AA, 8450\AA +                  & A-Band(O$_2$:7595\AA +), 7850-8450\AA   & \\
\hline
\multicolumn{3}{r}{TOTAL across all setups}   & 78406     & 8014\tablefootmark{d}                    &                                         &   \\
\hline
\end{tabular}
\tablefoot{
\tablefoottext{a}{Number of ESO archive spectra per setup, where RED L and RED U are counted separately.} \tablefoottext{b}{Approximate number of targets within coordinate search radius of 1.8'' per setup.} \tablefoottext{c}{Regions of Telluric Contamination (Tell.Cont.) measured in relative flux \tablefoottext{d}{ Approximate number of targets across all setups accounting for stars observed in multiple setups.}}
}
%RED564 & 4583-6686     & 5644-5654 & 5795      &  2467     & 5000-5130, 5370-5520, 5660-5750 & 5800-6000, 6270-6350, 6450-6610 & H$_{\alpha}$,H$_{\beta}$\\
%RED580 & 4726-6835     & 5804-5817 & 13958     &  3309     & 5000-5130, 5370-5520, 5660-5750 & 5800-6000, 6270-6350, 6450-6610 & H$_{\alpha}$,H$_{\beta}$ \\
%RED860 & 6650-10606    & 8544-8646 & 6734      &  1856     &              &  B Band (H$_2$O), A Band, H2 & Ca IR Triplet\\

\end{center}
%}
\end{table*}

The structure of this paper is as follows: Section~\ref{sec:AMBRE} reviews the AMBRE Project analysis in the context of the UVES spectra; Section~\ref{sec:uves_charac} characterises the UVES sample in terms of key measurables; Section~\ref{sec:rejcri} presents the rejection criteria identified for this sample and their application; Section~\ref{sec:calibsamps} presents the definition and application of the validation and calibration samples used in the analysis; Section~\ref{sec:intexterrors} presents the derivation of the internal and external errors; Section~\ref{sec:intersetupcomp} presents the inter-setup parameter comparision; Section~\ref{sec:discussion} presents the final stellar parameter results for UVES; and Section~\ref{sec:conclusion} concludes the paper.

\section{The AMBRE:UVES Stellar Parameterisation}\label{sec:AMBRE}

The stellar parameters that are determined in the AMBRE analysis are the effective temperature ($T_{\textrm{eff}}$ in K), surface gravity ($\log g$ in dex, where $g$ is in cm/s$^2$), mean metallicity ([M/H] in dex) and the enrichment in $\alpha$-elements with respect to iron ([$\alpha$/Fe] in dex). Here [M/H] is the global metallicity inferred from all elements heavier than He, not just Fe. Also we assume that the following chemical species are $\alpha$-elements: O, Ne, Mg, Si, S, Ar, Ca and Ti, although for any of the selected wavelength regions spectral features for all of these elements may not necessarily be present.

The synthetic grid of non-rotating FGKM-type spectra upon which the MATISSE analysis is carried out is described in detail in \citet{delaverny2012}. In summary this is a high resolution optical domain synthetic spectra grid calculated from the MARCS stellar atmophere models \citep{Gustafsson2008}, VALD atomic linelists \citep{Kupka1999} and molecular linelists provided by B.~Plez. The microturbulence ($\xi$) was hardwired into the grid such that for atmospheric models with high $\log g$ ($+3.5 \leq \log g \leq +5.5$) $\xi$ was set at 1.0~kms$^{-1}$ and for low $\log g$ ($\log g < +3.0$) $\xi$ was set at 2.0~kms$^{-1}$, these being typical values for dwarfs and giants respectively. 

The grid of 16783 flux normalized spectra covers the following ranges of atmospheric parameters: $T_{\textrm{eff}}$ between 2\,500~K and 8\,000~K, $\log g$ from $-0.5$ to $+5.5$~dex, and [Fe/H] from $-5.0$ to $+1.0$~dex, although not all combinations of the parameters are available within the grid. The selected MARCS models have [$\alpha$/Fe]=$0.0$ for [M/H] $\ge$ $0.0$, [$\alpha$/Fe]=$+0.4$ for [M/H] $\le$ $-1.0$ and, in between, [$\alpha$/Fe]=$-0.25$x[M/H]. For the spectra computation from each of these MARCS models, we considered an [$\alpha$/Fe] enrichment from $-0.4$ to $+0.4$~dex with respect to the canonical values that correspond to the original abundances of the MARCS models.

The AMBRE parameters ultimately reported to ESO lie within the parameter space of this grid, with some further restriction based on boundary issues. These restrictions are described further in Section~\ref{sec:rejparamspace}.

\subsection{UVES: Ultraviolet and Visual Echelle Spectrograph}\label{sec:UVES}
UVES is described in detail in \citet{Dekker2000}. It is a high resolution optical spectrograph used on the VLT and located at the Nasmyth B focus of UT2. Its wavelength coverage is from 3000~\AA\ to 11000~\AA. There are two arms for which the ultraviolet wavelengths are directed to the BLUE arm, and the visual wavelengths are directed to the RED arm. It has a resolving power of $\sim$40,000 when using a 1-arcsec slit. 

The BLUE arm covers a wavelength range from 3000 to 5000~\AA\ detected by a single CCD. The RED arm comprises two CCDs, lower (L) and upper (U), covering a wavelength range of 4200 to 11000~\AA. Standard templates with predefined central wavelengths for either or both of the arms are available as well as the ability to freely select the central wavelength for either or both arms.

For this project spectra from six standard setups were analysed: BLUE346, BLUE390, BLUE437, RED564, RED580 and RED860. The available wavelengths and number of spectra provided to OCA per setup are listed in Table~\ref{tab:setupinfo}. The BLUE setups were analysed separately from the RED setups even if objects were observed using the dichroic mode (BLUE+RED observed simultaneously). As a star may have been observed in multiple setups this results in the number of unique targets in each setup being much greater that the number of unique targets across all setups as seen in Table~\ref{tab:setupinfo}. 

For the RED setups the spectral total counts the L and U spectra separately but these were combined for the parameter analysis. The total spectra available are 78,406, while the total analysed, if L \& U are considered as one spectrum, are 51,921. Across all the setups this equates approximately to 8,014 distinct targets assuming a coordinate matching radius of 1.8'', the pointing accuracy of the VLT being $\sim 1$''. These distinct targets include a range of objects types such as supernova, quasars, variable stars, planets, hot stars and a variety of other exotic objects. There was no reduction of the sample to a pure stellar sample prior to receipt by OCA.

As well as operation in slit mode, UVES can also be fed via FLAMES for which 8 objects can be observed simultaneously. The archived spectra provided by ESO for the AMBRE Project are slit mode only.

\begin{figure}[!ht]
\centering
\begin{minipage}{90mm}
\vspace{-0.4cm}
%\hspace{-0.5cm}
\includegraphics[width=87mm]{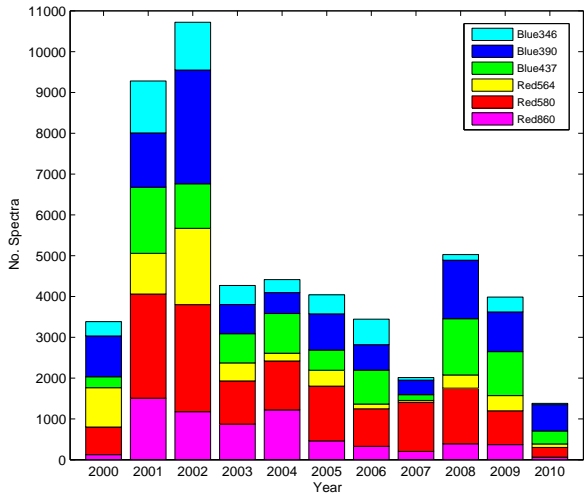}
%\vspace{-0.5cm}
\caption{Histogram of the number of spectra observed per year for each standard setup from March 2000 to November 2010.}\label{fig:setups_yearcounthist}
\end{minipage}
\end{figure}

The spectra delivered from ESO encompass observations from March 2000 to November 2010. Figure~\ref{fig:setups_yearcounthist} shows a histogram of the number of observations per year for the six standard setups considered here.

The spectra were reduced by ESO using the UVES reduction pipeline\footnote{www.eso.org/observing/dfo/quality/reproUVES/processing.html for the reduction steps}. Both single order and merged science-ready products were therefore available for the AMBRE:UVES analysis.

\subsection{AMBRE Pipeline Modifications}
As shown in Figure~4 of \citet{Worley2012}, the AMBRE pipeline comprises three stages, Spectral Processing A, B and C (referred to as SPA, SPB and SPC). SPA and SPB are the pre-processing stages for which the spectra are measured for radial velocity (V$_{rad}$) and spectral Full-Width-at-Half-Maximum (FWHM$_{spec}$), and problematic spectra are identified and rejected. SPC is the analysis stage where the final stellar parameters are determined by analysis of the spectra with MATISSE.

While the basic structure of the AMBRE pipeline remains the same, some aspects were improved for the UVES analysis. The measurement of the FWHM$_{spec}$ was extracted from SPB to be a standalone routine, like for V$_{rad}$. This proved to be more efficient as it could be run in parallel with the V$_{rad}$ procedure.

For FEROS the first guess of the stellar parameters from MATISSE provided in SPB was used for creating the set of synthetic spectra with which to carry out the first normalisation in SPC. This step was investigated further as it was observed that in some cases the rough normalisation used in SPB resulted in stellar parameters from MATISSE that returned quite an extreme parameter set and corresponding synthetic spectrum. This then sufficiently skewed the first normalisation of the observed spectrum in SPC that a valid solution was not recovered within the iterations of SPC.

Instead, using a standard synthetic spectrum for the first normalisation in SPC for all the spectra was found to retain such cases where a reasonable parameter determination was actually possible. Candidate synthetic spectra for this first SPC normalisation were investigated (for example the grid points closest to the Sun, Arcturus and Procyon). It was found that the synthetic spectrum of a cool metal poor dwarf was least likely to unncessarily skew the first normalisation of SPC thereby allowing the iterations to converge from a neutral starting point.

\subsection{UVES: Wavelength Configurations}
A key part of the AMBRE analysis is tailoring the pipeline configuration to the wavelength regions available. As there are six distinct setups to consider it was necessary to configure six different versions of the pipeline. 

This optimisation has three aspects: wavelength selection, resolution, and sampling. The most complex process was the optimal selection of the wavelength ranges where the goal was to minimise contamination by tellurics, avoid regions difficult to automatically normalise, and select wavelengths that were best represented by the synthetic spectra to be used in the analysis.

As for AMBRE:FEROS, we avoided using regions of significant telluric contamination when selecting the wavelengths for our analysis of the UVES spectra. As no telluric correction was carried out in the reduction by the ESO UVES pipeline, these features were present, and in some cases dominant, in some sections of the UVES spectra that were delivered to OCA. The degree of contamination was assessed by inspection of telluric atlases and are given as percentage of the relative flux in Table~\ref{tab:setupinfo}, whereby if the contribution to the flux of the spectrum by tellurics across a region was less than 5\% the region was considered uncontaminated. If the contribution was between 5\% and 20\% across a region it was considered as contaminated, and if above 20\% across the region it was considered strongly contaminated. A few key features noted within each setup are also listed. For the regions where there was between 5\% and 20\% contamination by telluric, these were only used if absolutely necessary, which was primarily the case for RED860 which suffers from extensive tellluric contamination. The regions with greater than 20\% contamination were rejected from consideration from the outset.

Some modifications to the routine were made based on the experiences with FEROS. For example optimising the sections of wavelength for normalisation such that: a) no region less than 20~\AA~ in extent was isolated from another region ($>$10~\AA~ distant); b) removing any very small isolated wavelength sections ($\sim$one pixel in extent) separated from adjoining regions by more than 0.5~\AA. These helped to avoid some normalisation difficulties encountered with the FEROS configuration.

All rejections were applied to a spectrum in its rest frame. Thus to account for the movement of tellurics from their observed wavelength when putting any spectrum in its rest frame, we included buffer regions at each end of any rejected region up to the maximum potential radial velocity shift equating to approximately 5\AA.

\begin{table*}[!htbp]
\tabcolsep=0.19cm
%TABLE Updated 22/9/2014
\caption{Final wavelength configuration for the MATISSE analysis for each UVES setup.}
\begin{center}
\begin{tabular}{lcccccccccccccc}
\hline\hline
 &  &  &  &  & \multicolumn{3}{c}{Convolution} &  & \multicolumn{3}{c}{Sun Flux Obs-Syn} & \multicolumn{3}{c}{Arcturus Flux Obs-Syn} \\ 
\hline
Setup & $\lambda_{min}$ & $\lambda_{max}$ & No. \AA\ & No. Pix & \AA/Pixel & {\scriptsize FWHM} (\AA) & R$_{min}$ & R$_{max}$ & $<$5\% & $<$10\% & $<$20\% & $<$5\% & $<$10\% & $<$20\% \\ 
BLUE346 & 3200 & 3850 & 527 & 8776 & 0.060 & 0.157 & 20000 & 24000 & 69 & 92 & 99 & 56 & 84 & 98 \\ 
BLUE390 & 3450 & 4400 & 604 & 6707 & 0.090 & 0.230 & 15000 & 19000 & 77 & 95 & 99 & 60 & 88 & 99 \\ 
BLUE437 & 3800 & 4950 & 967 & 13811 & 0.070 & 0.187 & 20000 & 26000 & 88 & 98 & 100 & 58 & 86 & 99 \\ 
RED564 & 4650 & 6450 & 850 & 9445 & 0.090 & 0.239 & 19000 & 27000 & 96 & 100 & 100 & 78 & 94 & 100 \\ 
RED580 & 4810 & 6750 & 505 & 5314 & 0.095 & 0.239 & 20000 & 28000 & 97 & 100 & 100 & 82 & 95 & 100 \\ 
RED860 & 6725 & 8900 & 322 & 2686 & 0.120 & 0.329 & 20000 & 27000 & 99 & 100 & 100 & 96 & 100 & 100 \\
\hline
\end{tabular}
\tablefoot{The adopted wavelength range, total number of angstroms and pixels, sampling, convolution FWHM, and resolution range are given. Also provided are the percentage of pixels for which the difference in flux between the Observed and the Synthetic is less than the limits of 5,10 and 20\%, for the comparison of the Sun and Arcturus Atlases to synthetic counterparts.}
\end{center}
\label{tab:setup_config}
\end{table*}

After this assessment, in order to select the final wavelength regions, a comparison was made between key spectral atlases and the corresponding synthetic spectra in order to identify those wavelength regions that are not well replicated by the synthetic models. This was carried out as a simple difference between atlas spectra and synthetic spectra, where the atlas spectra were convolved to match the expected resolution of the synthetic grid (R$\sim$20,000).

To identify gross discrepencies in spectral features between the observed and the synthetic spectra, Solar atlases and Arcturus atlases were used to identify feature mismatches. For identifying mismatches, the priority was given to the Sun, making the assumption that any gross discrepencies of the synthetic solar spectrum to the (convolved) atlas were a synthesis problem (incorrect/incomplete linelists or poorly modelled sections of synthetic spectra for example). Arcturus, while a well-known standard, is not as well studied as the Sun, so only very gross discrepencies were identified. Generally the \citet{SolarAtlasHinkle} and \citet{ArcturusAtlasHinkle} were used for each setup, except where the wavelength coverage of the setup exceeded that of the atlas, in which case alternate atlases were used \citep{Allende-Prieto2004,UVESPOP}.

Wavelengths about a spectral feature with flux differences between the solar atlas and the solar synthetic spectra (synthesised at the Sun stellar parameters) greater than 10\% were discarded. For the comparison of the Arcturus Atlas to the synthesised Arcturus spectrum, after the rejection based on the Sun, any further lines with a flux difference greater that 20\% were discarded.

Table~\ref{tab:setup_config} quantifies the resulting agreement of the final selected wavelengths with the atlases upon which the testing was carried out. This was carried out without use of the iterative normalisation process and simply shows how well the atlases and synthetic spectra agree with no optimisation of the normalisation. The first columns give the wavelength ranges, number of angstroms, number of pixels, convolution information and range of resolution across the wavelengths that was set for each UVES setup. The final six columns give the percentage of pixels for that setup that agreed between the atlas and the synthetic to better than 5, 10 and 20\% in terms of difference in flux. This empirical method resulted in the majority of the flux differences being less than 5\% for all six setups, particularly for the RED setups.

We particularly note here that such flux differences between observed and synthetic atlases should not strongly affect the stellar parameterisation. Indeed the MATISSE algorithm does not directly compare the observed and synthetic spectra but projects the observations on specific vectors containing information about how the flux varies as a function of the atmospheric parameters \citep[See][for further details]{Recio-Blanco2006,Bijaoui2008} .

The final two aspects for the training grid, resolution and sampling, were considered together. Once all possible wavelengths were rejected, with the goal of optimising the representation of the observed by the synthetic, the remainder for each setup was still significant in sampling and wavelength coverage. As for FEROS and HARPS, an optimisation of resolution and sampling was then made for each setup aiming to obtain a resolution of $\sim$20,000 (i.e. well below the original UVES spectral resolution). 

In particular, BLUE390 was the second setup tackled after RED580 (the two setups with the largest spectral sample). A similar convolution FWHM was adopted for BLUE390 as for RED580, which resulted in a lower resolution of the synthetic spectra grid as the wavelengths are much bluer. For the remaining setups a convolution FWHM was adopted that lead to a higher resolution of the respective synthetic grids. However the resolution adopted for the BLUE390 analysis is closer to that of the AMBRE:FEROS analysis, which is still more than high enough to retain the necessary spectral information required by MATISSE.

While degrading the spectra like this may seem a loss of essential information needed for parameterisation, several tests on this in previous studies revealed that for stellar parameterisation (but not for more precise chemical analysis) such a resolution is quite sufficient for robust determinations. See, for instance, some tests in \citet{Kordopatis2011} or the effect of the Gaia RVS resolution on the parametrisation in \citet{Recio-Blanco2016}.

\begin{figure}[!h]
\centering
\begin{minipage}{90mm}
\hspace{-0.4cm}
\includegraphics[width=95mm]{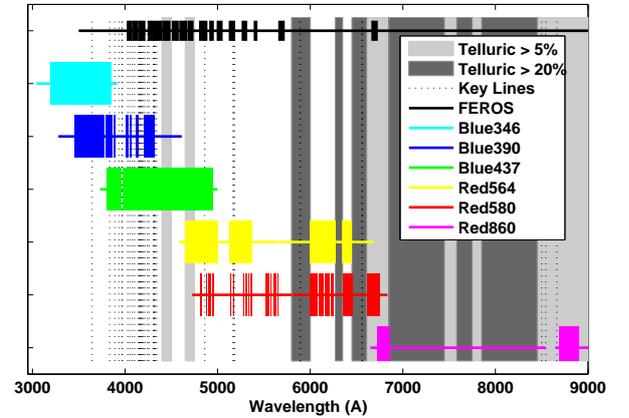}
\caption{Wavelength selection for the six UVES setups as provided in Table~\ref{tab:setup_config}. The selection for FEROS is shown for comparison. Key spectral features are indicated as dashed lines including: Hydrogen Balmer lines, Na, Ca, Mg, CN, C$_2$. }\label{fig:uves_wavecuts}
\end{minipage}
\end{figure}

The final configuration characteristics for each UVES setup are listed in Table~\ref{tab:setup_config} and the distribution of the wavelengths are shown in Figure~\ref{fig:uves_wavecuts}. 

%While the wavelength coverage for each UVES setup is not as extensive as the complete optical range for FEROS, each spectral domain still covers more angstroms at higher resolution than is required for a MATISSE analysis. The entire wavelength domain of each setup could be used for each configuration however this would involve significiant memory requirements. Instead, as for FEROS, an optimum configuration of number of pixels and resolution was determined.

\section{Characterisation of the UVES Spectra}\label{sec:uves_charac}
The dominant indicators for characterising the spectra are the S/N, the V$_{rad}$ and its associated errors ($\sigma_{\textrm{V}_{rad}}$) and the FWHM of the V$_{rad}$ cross-correlation function (FWHM$_{CCF}$). 

The S/N is calculated during the normalisation process as an estimate of the noise on the extracted pseudo continuum region used to normalise each section of spectra. Specifically it is the standard deviation of the ratio of the fluxes of the observed and synthetic spectra. However this type of estimate cannot be accurate in all cases, particularly, for example, across stellar type when the objects are cool and/or metal-rich. The many spectral features for particular wavelength regions result in an overestimation. Based on the iterative parameters we applied different clipping limits to try to reduce this effect. Other effects due to misfitting or spectral features and insturmental relics also influence the determination. An exact S/N determined during the ESO reduction process and provided wih the reduced spectra would be invaluable for automated parameterisation analyses.

The V$_{rad}$ is determined from the reduced UVES spectra using the method described in the AMBRE:FEROS analysis \citep{Worley2012}. In summary V$_{rad}$ is calculated by the cross-correlation of the spectrum to a set of binary masks. These masks were computed from the AMBRE synthetic spectra grid specifically for each of the six UVES setups analysed here.

The S/N and $\sigma_{\textrm{V}_{rad}}$ provide measures of the quality of the spectra. The FWHM$_{CCF}$ gives the first discriminator for spectra that cannot be analysed by the AMBRE Grid, identifying those spectra for which the spectral features are too broad for this grid of slow-rotating synthetic spectra. The rejection thresholds imposed for these measurements are described in Section~\ref{sec:preparameterrej}.

\begin{figure*}[!ht]
%\centering
\begin{minipage}{180mm}
%\vspace{-0.2cm}
\hspace{-1.0cm}
\includegraphics[width=195mm]{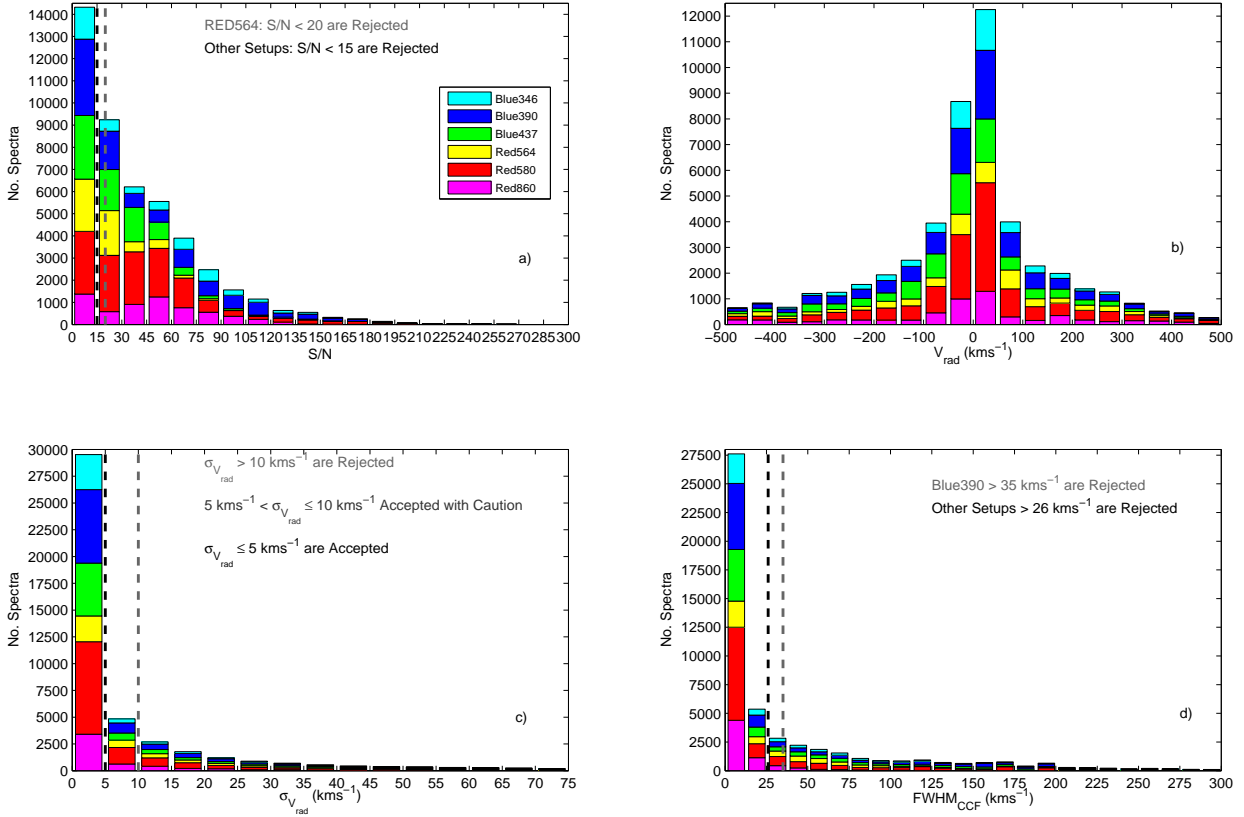}
\vspace{-1.3cm}
\caption{Stacked histograms of the number of spectra per Standard Setup as labelled for: a) S/N with rejection limit; b) V$_{rad}$; c) $\sigma_{\textrm{V}_{rad}}$ with upper rejection limit and intermediate threshold; d) FWHM$_{CCF}$ with rejection limit for BLUE390 and separately the other Setups.}\label{fig:setup_hists}
\end{minipage}
\end{figure*}

Figure~\ref{fig:setup_hists}a to d show the stacked histograms for these four indicators for each of the six standard setups. The key thresholds and how they are applied for S/N, $\sigma_{\textrm{V}_{rad}}$ and FWHM$_{CCF}$ are also shown.

The distribution of V$_{rad}$ in Figure~\ref{fig:setup_hists}b shows that approximately 60\% of the sample have a V$_{rad}$ between --100 and 100~kms$^{-1}$. This adheres to a general observational bias towards stars moving in a manner consistent with the galactic rotation. Looking at the outliers, 6\% of the total sample have an absolute value of V$_{rad}$ greater than 400~kms$^{-1}$ which is higher than expected ($<0.5\%$) in comparison to galactic V$_{rad}$ surveys such as RAVE \citep[][DR4]{Kordopatis2013}. However, only 0.6\% of the total sample have an absolute value of V$_{rad}$ greater than 400~kms$^{-1}$ and also have reported AMBRE parameters. Thus we have rejected the bulk (90\%) of the high V$_{rad}$ spectra within the parameter analysis. While there is no clear reason from the V$_{rad}$ quality assessment for these high V$_{rad}$ spectra it is likely that the V$_{rad}$ is indeed poorly measured for these spectra, evidenced by the lack of reported parameters. Those high velocity spectra  with reported parameters that remain in the final sample should be considered cautiously.

\section{Rejection Criteria}\label{sec:rejcri}
The AMBRE:FEROS analysis dealt extensively with characterising the types of spectra in the archive dataset and defining criteria that would identify any spectrum that could not be analysed by the pipeline. The criteria used in AMBRE:UVES are essentially the same as for AMBRE:FEROS with some variations in the thresholds depending on the wavelength configurations used for each setup. There are two sets of rejection criteria: Pre-Parameterisation (Pre-PM) whereby spectra that failed Pre-PM criteria would not have reliable parameters retrieved to due the nature of the spectra themselves; and Post-Parameterisation (Post-PM) for which the parameterisation itsself indicates that the derived parameters are not reliable.

The various criteria and associated thresholds used to assess and reject spectra are as follows:
\begin{enumerate}
 \item Pre-Parameterisation rejection criteria:
  \begin{enumerate}
    \item S/N below lower limit;
    \item $\sigma_{\textrm{V}_{rad}} >$ 10~kms$^{-1}$; 
    \item V$_{rad}$ CCF with negative contrast;
    \item $\frac{\sigma_{Amp}}{Amp} > 0.20$ and $\frac{\sigma_{Cont}}{Cont} > 0.10$;
    \item FWHM$_{spec}$ of medium strength spectral lines exceeds upper limit;
    \item FWHM$_{CCF}$ exceeds upper limit; 
  \end{enumerate}
 \item Post-Parameterisation rejection criteria:
  \begin{enumerate}
  \item Outside derived $\log\chi^2$-S/N relation limit;
  \item Parameters outside limits of synthetic spectra grid;
  \end{enumerate}
\end{enumerate}

where {\it Amp} is the amplitude of the CCF, and {\it Cont} is the location of the continuum of the CCF. See \citet{Worley2012} for more details.

\subsection{Pre-PM Rejections}\label{sec:preparameterrej}
As for AMBRE:FEROS, all spectra possible were put through the entire pipeline. In the first application of the spectral processing \citep[SPA, see Figure~4 of ][]{Worley2012} spectra that could not be analysed for reasons such as excessive noise, poor normalisation, instrumental artifacts, and extreme emission features (considered together as `Problematic' spectra) were discarded.

Rejection thresholds relating to S/N, $\sigma_{\textrm{V}_{rad}}$, FWHM$_{CCF}$ and FWHM$_{spec}$ are listed in Table~\ref{tab:rejcritthres} and explained below.
\begin{table}[!h]
%\vspace{-0.3cm}
\begin{center}
\caption{Rejection criteria thresholds for each UVES setup.}\label{tab:rejcritthres}
\begin{tabular}{lcccc}
\hline\hline
UVES  & S/N\tablefootmark{a} & $\sigma_{\textrm{V}_{rad}}$\tablefootmark{b} & FWHM$_{CCF}$\tablefootmark{b}  & FWHM$_{spec}$\tablefootmark{b} \\
Setup &     & (kms$^{-1}$)  & (kms$^{-1}$)            &  (\AA) \\
\hline
BLUE346 & 15 & 10 &  26 & 0.20 \\
BLUE390 & 15 & 10 & 35 & 0.30 \\
BLUE437 & 15 & 10 & 26 & 0.25 \\
RED564 & 20 & 10 & 26 & 0.30 \\
RED580 & 15 & 10 & 26 & 0.30 \\
RED860 & 15 & 10 & 26 & 0.40 \\
\hline
\end{tabular}
\tablefoot{
\tablefoottext{a}{Lower limit.}\tablefoottext{a}{Upper limit.}
}
\end{center}
%\vspace{-0.3cm}
\end{table}

The S/N was measured at each stage of the analysis as in the AMBRE:FEROS analysis with the reported S/N being that calculated using the synthetic spectrum at the final accepted parameters thus being the best estimate of the S/N. Spectra with too low S/N do not have sufficient signal with which to derive reliable parameters. A threshold of 15, as shown in Figure~\ref{fig:setup_hists}a was adopted for the AMBRE:UVES analysis, except for RED564 for which a threshold of 20 was adopted due to obvious outliers identified by visual inspection. These thresholds are slightly larger than that adopted in previous AMBRE analyses \citep[See][]{Worley2012,DePascale2014}. This is because the present UVES spectra cover a much smaller wavelength domain per setup than those of FEROS and HARPS. In consequence, fewer spectral signatures may be available for the parameterisation and their obscuration has a greater impact on the reliability of parameters at a higher level of S/N. Hence this stricter selection on the spectra quality. Approximately 23.1\% of the spectra across the setups have a S/N less than 15 (less than 20 for RED564).

The upper limit on the $\sigma_{\textrm{V}_{rad}}$ is the same as was used for the previous AMBRE analyses as the radial velocity programme is unchanged and the spectra at the observed high resolution (not convolved to the grid resolution) were used for the radial velocity determination. Approximately 28.4\% of the analysed UVES spectra have a $\sigma_{\textrm{V}_{rad}}$ greater than 10~kms$^{-1}$.

The FWHM$_{CCF}$ gives an indication of the spectral broadening, where a high FWHM$_{CCF}$ is typical of hot and/or fast rotating stars, as well also possibly indicating a mismatch of the spectrum with the binary mask which is often accompanied by irregular values in the other $\sigma_{\textrm{V}_{rad}}$ quality indicators. Those values remaining after rejection we considered to be representative of the astrophysical broadening. \citet{Gazzano2010} provides limits on the CCF, for the resolution of a particular synthetic grid, above which the spectral features are too broad for parameters to be reliably determined, i.e. astrophysical broadening greater than the limit is no longer masked by the convolution of the grid. As for AMBRE:FEROS, for each UVES configuration the relevant threshold was extrapolated from \citet{Gazzano2010}. There is some variation between setups based on the resolution and sampling optimisation (See Table~\ref{tab:rejcritthres}) and across the entire sample approximately 38.3\% have a FWHM$_{CCF}$ greater than the respective setup threshold.  

The width of the spectral features themselves (FWHM$_{spec}$) was also used to reject spectra, where thresholds were defined empirically by relating FWHM$_{spec}$ to FWHM$_{CCF}$ via $\sigma_{T_{\textrm{eff}}}$ as for AMBRE:FEROS \citep[See][]{Worley2012}.

The total number of spectra rejected due to Pre-PM criteria are listed in Table~\ref{tab:rejcountsprespc}. A spectrum may be rejected based on several criteria, hence the sum of the rejections as listed is greater than the final rejected total. For Table~\ref{tab:rejcountsprespc} a RED setup spectrum is considered as the L+U arms combined as per the analysis process and thus the spectral totals are different to the totals in Table~\ref{tab:setupinfo}. Table~\ref{tab:rejcountsprespc} includes the spectral totals of: the initial spectra per setup; too low S/N, poor template matching based on the $\textrm{V}_{rad}$ analysis, too large FWHM$_{CCF}$, too large $\textrm{V}_{rad}$ error, too large broadening (FWHM$_{spec}$) of the spectral features, problematic spectra as mentioned above, and the total finally rejected by these Pre-PM criteria. Rejected spectra were mainly too low in S/N, too broad in V$_{rad}$ CCF or too high in $\sigma_{\textrm{V}_{rad}}$, and often failed on more than one criteria.

\begin{table*}[!ht]
\begin{center}
\tabcolsep=0.25cm
\caption{Total spectra and pre-PM rejections per UVES setup.}
\begin{tabular}{lcccccccc}
\hline\hline
Setup  &   \# Spectra\tablefootmark{a}     &  S/N       & $\textrm{V}_{rad}$  & FWHM$_{CCF}$ &  $\sigma_{\textrm{V}_{rad}}$  & FWHM$_{spec}$  &  Problematic        &     Total Rejected  \\
       &                  &  $<$ Thres.     & Fit Error           & $>$ Thres.   &   $>$ 10 kms$^{-1}$           &    $>$ Thres.   &   Spectra             &     Pre-PM  \\
\hline
BLUE346   &  5267   &    838     &              162               &             2383              &              1446                   &          1093        &          184     &   2980       \\
BLUE390   &  11262   &    3454     &              1681               &             4207              &              3034                   &          333        &          471     &   7704       \\
BLUE437   &  8907   &    2914     &              1048               &             3874              &              2795                   &          1557        &          619     &   6264       \\
RED564   &  5795   &    1101     &              1416               &             2891              &              2331                   &          270        &          679     &   4555       \\
RED580   &  13956   &    2202     &              137               &             4959              &              3482                   &          2132        &          644     &   7257       \\
RED860   &  6734   &    1476     &              148               &             1595              &              1664                   &          379        &          1211     &   3546       \\
Total   &  51921   &    11985     &              4592               &             19909              &              14752                   &          5764        &          3808     &   32306       \\
\hline
\% of Total   &  100.0   &    23.1     &              8.8               &             38.3              &              28.4                   &          11.1        &          7.3     &   62.2       \\
\hline
\end{tabular}
\tablefoot{
\tablefoottext{a}{RED L+U spectra are considered as one spectrum.}
}
\label{tab:rejcountsprespc}
\end{center}
\end{table*}

\subsection{Post-PM 1: S/N \& $\log(\chi^2)$}\label{sec:snrlchi2}
The majority of the total UVES dataset were put through the entire parameterisation process. The Pre-PM criteria were then applied to obtain the preliminary cleaned sample (19615 spectra) that was then assessed by the two Post-PM rejection criteria.

\begin{figure*}[!ht]
\centering
\begin{minipage}{180mm}
\vspace{1cm}
\hspace{-2.0cm}
\includegraphics[width=205mm]{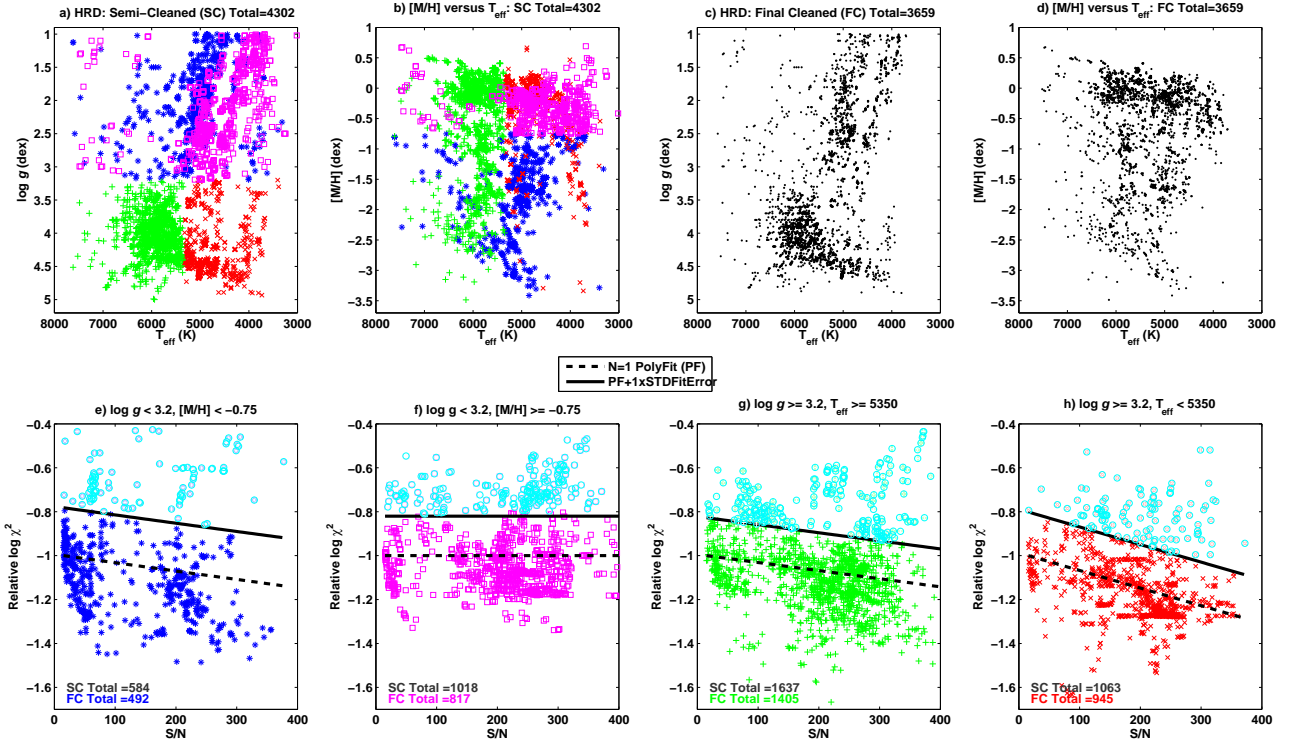}
\vspace{-0.8cm}
\caption{Diagnostics for cleaning the RED580 sample using S/N and $\log(\chi^2)$: a) HR Diagram color coded to four subsamples; b) $T_{\textrm{eff}}$ vs [M/H] showing four subsamples; c) Cleaned HR Diagram; d) Cleaned $T_{\textrm{eff}}$ vs [M/H]; e) S/N vs $\log(\chi^2)$ for metal-poor giant subsample (blue \textcolor{blue}{*}); f) S/N vs $\log(\chi^2)$ for metal-rich giant subsample (magenta \textcolor{magenta}{$\Box$});   g) S/N vs $\log(\chi^2)$ for hot dwarf subsample (green \textcolor{green}{+}); h) S/N vs $\log(\chi^2)$ for cool dwarf subsample (red \textcolor{red}{x}); . Rejections are cyan \textcolor{cyan}{o} and final accepted spectra are black $\bullet$.}\label{fig:logchi_red580}
\vspace{1cm}
\end{minipage}
\end{figure*}

The first of these is the S/N-$\log(\chi^2)$ relation. The $\log(\chi^2)$ is a measure of the agreement between the observed spectrum and the synthetic spectrum which has been calculated at the corresponding MATISSE stellar parameters. As the noise decreases (S/N increases) the estimate of the stellar parameters is improved, thus the synthetic spectrum is a better fit. In theory as the difference between the observed and synthetic goes to zero so too does $\chi^2$ go to zero (and $\log(\chi^2)$ goes to negative infinity). However, there will never be an exact match between the observed and synthetic (until at least we can model stellar spectra perfectly). Hence $\chi^2$ will never be exactly zero, and there will be a lower limit to $\log(\chi^2)$ with increasing S/N. This lower limit is evident in Figure~\ref{fig:logchi_red580}b, c, f \& g although the exponential decay-like trend is better seen in the HARPS sample in \citet{DePascale2014}, as the UVES samples are unevenly distributed across the S/N range.

Those observed spectra for which the derived parameters are not well-defined will not adhere to this trend lying as scatter well above the limit. The exact placement of the relation also differs between wavelength ranges and spectral types. Therefore the S/N and $\log(\chi^2)$ distribution was explored for each setup to derive relations by which outliers could be rejected. For the majority of the setups, separating values into subsamples according to $\log g$, $T_{\textrm{eff}}$ and/or [M/H] showed a clearer distribution for the different sets of stars.

\begin{table*}[!ht]
\centering
\caption{Parameter space thresholds and rejection relation coefficients per UVES setup.}
\begin{tabular}{lccccclccc}
\hline\hline
 & No. & \multicolumn{ 2}{c}{Thresholds} & Dwarf & Giant &  & \multicolumn{ 2}{c}{Linear Coefficients} & \multicolumn{1}{c}{$<$Fit Error$>$} \\ 
 & Subsamples & S/N & $\log g$ & $T_{\textrm{eff}}$ & [M/H] & Subsample & \multicolumn{1}{c}{{\bf a}} & \multicolumn{1}{c}{{\bf b}} & \multicolumn{1}{c}{{\bf FE}} \\ 
\hline
 BLUE346 & 3 & 15 & 3.0 & 6000 &  & Giants  & -0.0036 & -1.1032 & 0.5385 \\ 
 &  &  &  &  &  & Hot Dwarfs  & -0.00007 & -2.2019 & 0.3511 \\ 
 &  &  &  &  &  & Cool Dwarfs  & -0.0023 & -1.4063 & 0.2937 \\ 
BLUE390 & 3 & 15 & 3.5 & 5700 &  & Giants  & -0.0068 & -1.4778 & 1.0669 \\ 
 &  &  &  &  &  & Hot Dwarfs  & -0.0016 & -1.9162 & 0.4333 \\ 
 &  &  &  &  &  & Cool Dwarf  & -0.0046 & -1.1529 & 0.5155 \\ 
BLUE437 & 3 & 15 & 3.7 &  & -1.50 & Metal-Poor Giants  & -0.0028 & -2.4660 & 0.4890 \\ 
 &  &  &  &  &  & Metal-Rich Giants  & -0.0018 & -1.2019 & 0.4071 \\ 
 &  &  &  &  &  & Dwarfs  & -0.0017 & -2.1013 & 0.4743 \\ 
RED564 & 1 & 20 &  &  &  & All  & -0.0084 & -1.7858 & 0.5903 \\ 
RED580 & 4 & 15 & 3.2 & 5350 & -0.75 & Metal-Poor Giants  & -0.0011 & -2.8774 & 0.6310 \\ 
 &  &  &  &  &  & Metal-Rich Giants  & 0.0002 & -2.4746 & 0.4440 \\ 
 &  &  &  &  &  & Hot Dwarfs  & -0.0011 & -2.9635 & 0.5142 \\ 
 &  &  &  &  &  & Cool Dwarfs  & -0.0020 & -2.4642 & 0.4948 \\ 
RED860 & 3 & 15 & 4 & 5600 &  & Giants  & 0.0000 & -3.0721 & 0.5406 \\ 
 &  &  &  &  &  & Hot Dwarfs  & -0.0021 & -2.8524 & 0.5080 \\ 
 &  &  &  &  &  & Cool Dwarfs  & -0.0033 & -2.9257 & 0.6039 \\ 
\hline
\end{tabular}
\tablefoot{The parameter thresholds are used to define subsamples within each setup for defining corresponding rejection relations. Results are rejected based on the linear equation: $\log(\chi^2)_{rej} >$~{\bf a}S/N~+~{\bf b}~+~{\bf FE}.}
\label{tab:logchi_threscoeffs}
\vspace{1cm}
\end{table*}

The separation between dwarfs and giants, hot dwarfs and cool dwarfs, metal-rich giants and metal-poor giants were defined empirically from inspection of the HR diagram and temperature versus metallicity distribution for each setup. The thresholds used to define the subsamples per setup are listed in Table~\ref{tab:logchi_threscoeffs}.

\begin{table*}[!ht]
\centering
\caption{Totals for Rejected (Rej) and Accepted (Acc) spectra for the Post-PM samples.}
\begin{tabular}{lccccccccc}
\hline\hline
UVES       &  Total        &  Rej.     &  Rej.    & Total Rej.       &  Total Acc.\tablefootmark{b}    & Total Acc.    & Total Acc.    & Total       &  Approx. \\ 
Setup      &  Post-PM      &  S/N-LC\tablefootmark{a}        &   POG\tablefootmark{b}        &  Post-PM             &  TON               &  TGM              &   TGMA            & Acc.    &  Stars   \\ 
\hline
BLUE346     &   2287        &    458         &   154        &   612                &  92               &   80              &   1503             &    1675      &    628   \\ 
BLUE390     &   3558        &    539         &   1334        &   1873                &  322               &   45              &   1318             &    1685      &    882   \\ 
BLUE437     &   2643        &    802         &   199        &   1001                &  168               &   60              &   1414             &    1642      &    878   \\ 
RED564     &   1240        &    35         &   448        &   483                &  174               &   33              &   550             &    757      &    382   \\ 
RED580     &   6699        &    1555         &   421        &   2117                &  810               &   99              &   3673             &    4582      &    1462   \\ 
RED860     &   3188        &    774         &   352        &   1126                &  187               &   121              &   1754             &    2062      &    697   \\ 
\hline
Total     &   19615        &    4163         &   2908        &   7212                &  1753               &   438              &   10212             &    12403      &    3708   \\ 
\% Post-PM     &   100.0        &    21.2         &   14.8        &   36.8                &  8.9               &   2.2              &   52.1             &    63.2      &    70.4   \\ 
\% Total     &   37.8        &    8.0         &   5.6        &   13.9                &  3.4               &   0.8              &   19.7             &    23.9      &    46.2   \\ 
\hline
\end{tabular}
\tablefoot{
\tablefoottext{a}{S/N-$\log(\chi^2)$ relation.}
\tablefoottext{b}{Parameters Outside Grid.}
}
\label{tab:postpm_cleaned}
\end{table*}

The fitting was inspected visually for each setup and subsample, and an upper limit of the linear fit plus the addition of the mean fit error, rather than some multiple thereof, was the most appropriate in all cases. The cyan circles are those spectra that were rejected as being indicative of poor fits, including those with high S/N. An example of such a case may be a spectroscopic binary for which the observation has good signal but the dual spectral features would result in a poor fit to the (erroneous) solution spectrum.

The coefficients ({\bf a},{\bf b}) and mean fit error ({\bf FE}) are listed in Table~\ref{tab:logchi_threscoeffs}. {\bf FE} is the output from the polynomial fitting routine and is defined as an estimate of the standard deviation of the error in predicting a future observation at x by p(x). For each setup and subsample, spectra were rejected if they satisfied the following relation: $\log(\chi^2) >$~{\bf a}S/N~+~{\bf b}~+~{\bf FE}.

Figure~\ref{fig:logchi_red580} illustrates the process of defining the S/N-$\log(\chi^2)$ relations for the RED580 spectra as an example. For better comparison within the figure each sample is normalised in $\log(\chi^2)$ such that the linear fit at the minimum S/N has a $\log(\chi^2)$ of -1. This we have defined as the `Relative $\log(\chi^2)$' and it shows how the gradient of the linear relation (dashed black line) differs between samples. The lower limit of rejection in each sample was set by the linear relation plus {\bf FE} (solid black line).

\subsection{Post-PM 2: Grid Parameter Space}\label{sec:rejparamspace}
The final stage of rejection is to impose criteria based on the boundaries of the synthetic grid parameter space. These are spectra which have satisfied all quality and suitability criteria but the parameters have been derived sufficiently close to (or outside) the grid boundaries such that the solutions may not be well-formed. As for AMBRE:FEROS and AMBRE:HARPS\footnote{Lower $T_{\textrm{eff}}$ limit is 4000~K for AMBRE:HARPS}, the boundaries of the grid for assuming reliable results are:

\begin{description}
 \item $\hspace{1.0cm}3000 \leq T_{\textrm{eff}} \leq 7625 $~K
 \item $\hspace{1.0cm}\hspace{0.2cm}1.0 \leq \log g \leq 5.0 $~dex ([$g$]=cm/s$^2$)
 \item $\hspace{1.0cm}-3.5 \leq [\textrm{M/H}] \leq 1.0 $~dex
 \item $\hspace{1.0cm}-0.4 \leq [\alpha/\textrm{Fe}] \leq 0.4$~dex$    \hspace{0.6cm} \textrm{if  } [\textrm{M/H}] \geq 0.0$
 \item $\hspace{1.0cm}-0.4 \leq [\alpha/\textrm{Fe}] \leq 0.8$~dex$   \hspace{0.6cm} \textrm{if  } -1.0 < [\textrm{M/H}] < 0.0$
 \item $\hspace{1.0cm} \hspace{0.2cm} 0.0 \leq [\alpha/\textrm{Fe}] \leq 0.8$~dex$    \hspace{0.6cm}  \textrm{if  } [\textrm{M/H}] \leq -1.0$
\end{description}

The final count for the cleaned sample of UVES spectra per setup are listed in Table~\ref{tab:postpm_cleaned}. The number of spectra rejected based on the S/N-$\log(\chi^2)$ relations (S/N-LC), and those rejected with Parameters Outside the Grid (POG) are given, then the total number of rejections based on these two criteria. As for AMBRE:FEROS, three categories of accepted parameters were defined:

\begin{enumerate}
 \item TON: T$_{\textrm{eff}}$ only is accepted within the grid parameter limits,
 \item TGM: T$_{\textrm{eff}}$, $\log g$, and [M/H] are accepted,
 \item TGMA: T$_{\textrm{eff}}$, $\log g$, [M/H] and [$\alpha$/Fe] are accepted.
\end{enumerate}

The total number of spectra accepted for each of these categories are given in Table~\ref{tab:postpm_cleaned} as well as their percentages with respect to the total spectra assessed Post-PM, and their percentages with repsect to the total UVES spectral sample. This final cleaned sample with accepted parameters (12403 spectra) was used for the following calibration and error analyses.

\section{Validation \& Calibration Samples}\label{sec:calibsamps}
The pipeline configuration for each setup needed to be calibrated and the results validated. As for AMBRE:FEROS, stellar atlases and samples of stars from the PASTEL database \citep[][2013 update available on VizieR\footnote{http://vizier.u-strasbg.fr/viz-bin/VizieR}]{Soubiran2010}  were used for this process, together with a new set of calibration `Benchmark' stars.

Since the completion of the AMBRE:FEROS analysis considerable effort has been made within the stellar spectroscopic community to identify a standard set of stars with well-defined stellar parameters that provide reasonable coverage of the $T_{\textrm{eff}}$-$\log g$-[Fe/H] parameter space. This sample of 34 Benchmark stars have been compiled primarily for the Gaia Mission but are already being used extensively in the Gaia-ESO Survey \citep[][and references therein]{Jofre2014}.

\begin{figure*}[!ht]
\centering
\begin{minipage}{180mm}
%\hspace{-1.3cm}
\includegraphics[width=180mm]{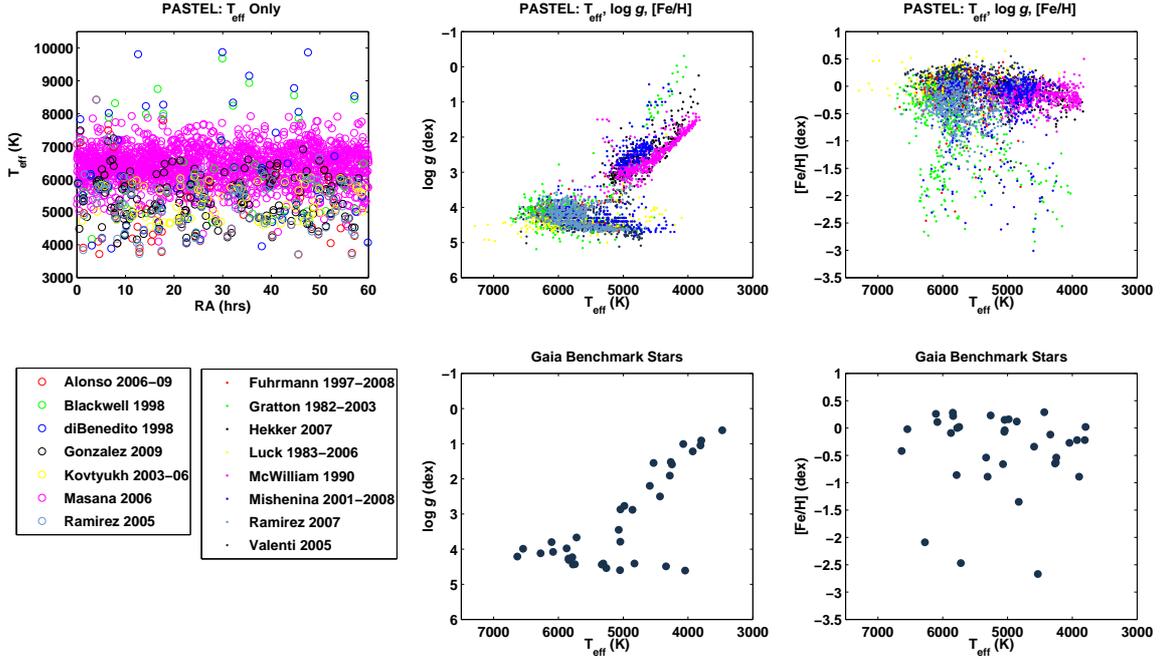}
\caption{Distribution of the PASTEL and Gaia Benchmark samples in parameter space: a) PASTEL $T_{\textrm{eff}}$-only papers for RA versus $T_{\textrm{eff}}$; b) H-R diagram of PASTEL papers reporting $T_{\textrm{eff}}$, $\log g$ and [Fe/H]; c) as for b) but $T_{\textrm{eff}}$ versus [Fe/H]; d) as for b) but for the Gaia Benchmark sample; e) as for c but for the Gaia Benchmarks.}\label{fig:pastel_gaiabm}
\end{minipage}
\end{figure*}

As this list was available for the UVES analysis it was possible to search for some subsample of the Benchmark stars within the spectra for each setup. Table~\ref{tab:summary_pgbma} lists the spectra found for each setup for each Benchmark star, the number of spectra in the PASTEL sample for each setup and the stellar atlases used for each UVES setup. Figure~\ref{fig:pastel_gaiabm} displays the PASTEL and Benchmark samples exploring their distribution in $T_{\textrm{eff}}$, $\log g$ and [Fe/H].

\begin{table}[htbp]
\tabcolsep=0.10cm
\caption{Atlases, Calibrators and associated UVES spectra per setup.}
\begin{center}
\begin{tabular}{c|cc|cc|ccc}
\hline\hline
 & BM & UVES  & Pastel & UVES & \multicolumn{3}{c}{Atlases}  \\ 
 & Stars & Spectra & Stars & Spectra & Sun & Arcturus & Procyon \\ 
\hline 
BLUE346 & 18 & 303 & 305 & 897 & 1 & ... & ... \\ 
BLUE390 & 4 & 29 & 190 & 746 & 3 & 2 & 1 \\ 
BLUE437 & 16 & 151 & 219 & 636 & 4 & 3 & 2 \\ 
RED564 & 3 & 7 & 65 & 209 & 4 & 3 & 2 \\ 
RED580 & 20 & 588 & 373 & 2283 & 4 & 3 & 2 \\ 
RED860 & 16 & 139 & 235 & 911 & 1 & 1 & 1 \\ 
\hline
\end{tabular}
\end{center}
\label{tab:summary_pgbma}
\end{table}

The two setups with greatest number of spectra are RED580 and BLUE390, followed by BLUE437, RED860, RED564 and BLUE346 in descending order. Each setup (wavelength range) needed to be treated individually with a tailored pipeline and MATISSE configuration. Multiple configurations were tested for each setup to explore the respective wavelength domains. The final configurations provided the most reliable results based on the calibration samples.

\begin{figure*}[!ht]
%\centering
\begin{minipage}{180mm}
\hspace{-2.2cm}
\includegraphics[width=220mm]{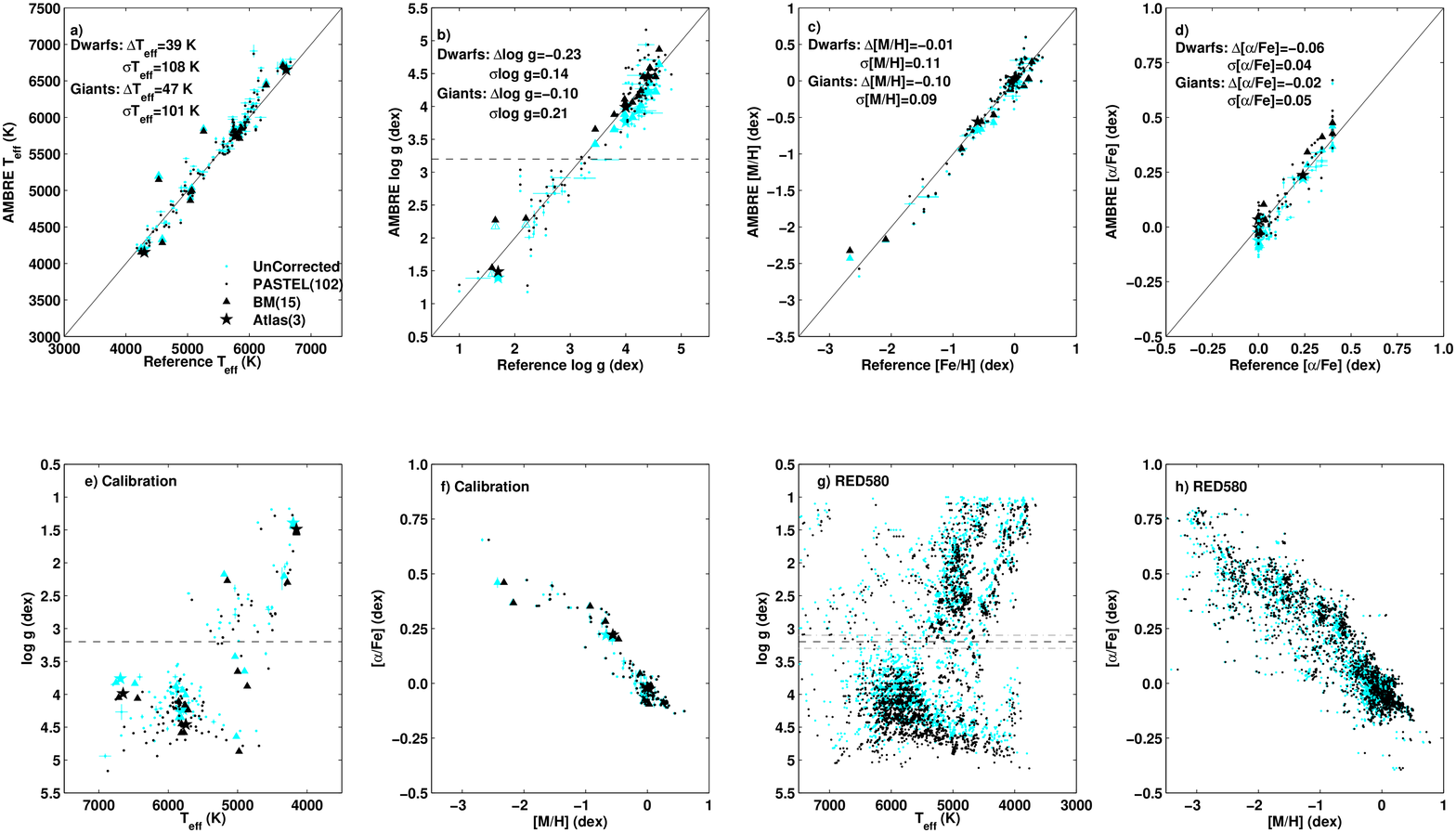}
\vspace{-1cm}
\caption{Comparison of AMBRE results of parameters determination for RED580 PASTEL, Benchmark and Atlas samples with accepted values. a) $T_{\textrm{eff}}$; b) $\log g$;  c) [Fe/H] vs [M/H]; d) [$\alpha$/Fe] vs $\alpha$-relation; e) HR Diagram - Calibration ; f) [M/H] vs [$\alpha$/Fe] - Calibration; d) RED580 HR Diagram; h) RED580 [M/H] vs [$\alpha$/Fe]. Raw results are in cyan, bias-corrected results are in black. The black dashed line shows the $\log g$ limit used to separate giants and dwarfs. The grey dot-dash lines indicate the crossover region}\label{fig:red580_calib}
\end{minipage}
\end{figure*}

As an example, Figure~\ref{fig:red580_calib} shows the calibration process for the RED580 configuration. The three key samples (PASTEL, Benchmarks, Atlases) were used to derive what, if any, bias corrections were needed to obtain agreement between the AMBRE parameters and the accepted parameters of those samples. 

First, this comparison was carried out per star rather than per spectrum in order to negate the effect of any outliers and to avoid the biases being weighted by multiple observations of the same star. Hence, where there were multiple spectra per star, or multiple PASTEL parameters per star, these were reduced to a mean parameter (AMBRE or PASTEL) with standard deviation. For any star with multiple spectra only those spectra per star were retained that fell within the following spread in parameters for either the AMBRE or Reference parameters:

\begin{itemize}
 \item $\sigma_{T_{\textrm{eff}}}<100$~K,
 \item $\sigma_{\log g}<0.25$~dex ([$g$]=cm/s$^2$),
 \item $\sigma_{[M/H]}<0.25$~dex   \hspace{0.5cm} if $[M/H]<-1.0$, or
 \item $\sigma_{[M/H]}<0.15$~dex   \hspace{0.5cm} if $[M/H]>-1.0$.
\end{itemize}

For the reference parameters this was particularly the case for the PASTEL sample, for which multiple studies analysed the same star. This ensured the cleanest comparison possible between the sets of results.

The corrections were then applied to the full sample per spectrum (not per star). In each sub-figure of Figure~\ref{fig:red580_calib}, the raw AMBRE:UVES:RED580 per spectrum results are shown as cyan points, while the results per spectrum corrected for the biases are shown as black points.

\begin{table*}[!ht]
\tabcolsep=0.15cm
\centering
\caption{Per setup bias and dispersion of AMBRE parameters compared to reference parameters.}
\begin{tabular}{lcccccccccccc}
\hline\hline
Setup  & SubSample & lim($T_{\textrm{eff}}$) & lim($\log g$) & $\Delta T_{\textrm{eff}}$ & $\sigma_{T_{\textrm{eff}}}$ & $\Delta \log g$ & $\sigma_{\log g}$ & $\Delta$[M/H] & $\sigma_{\textrm{[M/H]}}$ & $\Delta$[$\alpha$/Fe] & $\sigma_{[\alpha\textrm{/Fe}]}$ & No. Stars \\ 
\hline
BLUE346 & Dwarf & $>$6000 & $>$3.0 & 208 & 66 & -0.04 & 0.16 & -0.08 & 0.07 & -0.12 & 0.03 & 28 \\ 
 & Dwarf & $<$6000 & $>$3.0 & 4 & 149 & -0.22 & 0.21 & -0.30 & 0.14 & -0.18 & 0.05 & 47 \\ 
 & Giant &  & $\leq$3.0 & -55 & 105 & 0.00 & 0.23 & -0.53 & 0.12 & -0.13 & 0.09 & 20 \\ 
 &  &  & \multicolumn{1}{l}{} &  &  &  &  &  &  &  &  &  \\ 
BLUE390 & All &  & - & -124 & 105 & -0.42 & 0.23 & -0.31 & 0.09 & 0.04 & 0.06 & 51 \\ 
 &  &  &  &  &  &  &  &  &  &  &  &  \\ 
BLUE437 & Dwarf &  & $>$3.7 & 82 & 92 & 0.05 & 0.20 & -0.09 & 0.06 & -0.01 & 0.04 & 40 \\ 
 & Giant &  & $\leq$3.7 & 190 & 111 & 0.29 & 0.30 & 0.03 & 0.17 & -0.04 & 0.05 & 27 \\ 
 &  &  &  &  &  &  &  &  &  &  &  &  \\ 
RED564 & All &  & - & -128 & 80 & -0.36 & 0.13 & -0.16 & 0.08 & 0.02 & 0.07 & 16 \\ 
 &  &  & \multicolumn{1}{l}{} &  &  &  &  &  &  &  &  &  \\ 
RED580 & Dwarf &  & $>$3.2 & 39 & 108 & -0.23 & 0.14 & -0.01 & 0.11 & -0.06 & 0.04 & 88 \\ 
 & Giant &  & $\leq$3.2 & 47 & 101 & -0.10 & 0.21 & -0.10 & 0.09 & -0.02 & 0.05 & 29 \\ 
 &  &  &  &  &  &  &  &  &  &  &  &  \\ 
RED860 & Dwarf &  & $>$4.0 & -164 & 124 & 0.18 & 0.19 & -0.06 & 0.21 & -0.04 & 0.06 & 33 \\ 
 & Giant &  & $\leq$4.0 & -78 & 198 & -0.3$x\log g+$1.3 & 0.25 & 0.05 & 0.28 & -0.03 & 0.06 & 32 \\ 
\hline
\end{tabular}
\tablefoot{Subsamples are defined as Dwarf and Giant with respective gravity and temperature limits defined. Sigma values used as reported external error (comparison to external source).}
\label{tab:biases}
\end{table*}

Figure~\ref{fig:red580_calib}a, b, c \& d show the Reference parameter to AMBRE parameter per star comparison for $T_{\textrm{eff}}$, $\log g$, [M/H] and [$\alpha$/Fe].
The mean difference (bias) and standard deviation (external error) between the parameters for the dwarf sub-sample and for the giant sub-sample are included and also provided in Table~\ref{tab:biases}. Figure~\ref{fig:red580_calib}e \& f show the calibration sample HR diagram and the comparison of the [M/H] with [$\alpha$/Fe] respectively. Figure~\ref{fig:red580_calib}g \& h are as for e \& f but for the full RED580 sample.

For the $\log g$ plots, the black dashed line is the limit used to separate the dwarf and giant samples, which corresponds to the $\log g$ threshold set for the S/N-$\log(\chi^2)$ investigation. In the RED580 HR Diagram the grey dot-dashed lines define the cross-over region (0.2 dex in extent) in which each parameter correction is a linear interpolation between the parameter offsets of the two sub-samples.

Biases were also calculated for [$\alpha$/Fe] by comparison of the AMBRE values against the [$\alpha$/Fe] relation that has been hard-wired into the synthetic grid \citep{delaverny2012}, whereby the accepted [Fe/H] was used to estimate an expected [$\alpha$/Fe]. However these biases (see Table~\ref{tab:biases}) were not applied as corrections because they cannot be independently verified. 

Overall there is good agreement between the AMBRE parameters and the expected parameters for each sample. The main discrepency in this setup is the determination of the $\log g$ for the dwarf sample with a 0.2~dex offset between the AMBRE value and the accepted values. For the giants the spread in $\log g$ is much greater than that for the dwarfs but the bias is less than half that of the dwarf sample. 

Table~\ref{tab:biases} provides the mean difference and standard deviation for each sub-sample for each parameter for each of the six UVES setups. These were applied to the cleaned per spectrum sample to correct the parameters to their final reported values. We re-emphasise that the [$\alpha$/Fe] biases are shown but not applied to the sample. The limit at which each sample was separated into dwarfs and giants is also given. The cross-over region of $\pm 0.1$~dex was defined about this limit in each case to provide a smooth linear transition between the two sub-samples in the parameter corrections. Specific codicils are:

\begin{itemize}

\item BLUE346 $\rightarrow$ it was necessary to further divide the dwarf sample by a temperature limit as the hot end of the dwarf sample skewed the offset of the bulk of that sample. For the lower temperature dwarfs a relatively large bias in $\log g$ and in [M/H] was found, although a very small offset in $T_{\textrm{eff}}$;

\item BLUE390 $\rightarrow$ after cleaning, the giant calibration sample was limited to a single Arcturus atlas, hence determining biases separated by dwarfs and giants was dubious. Consequently the biases are based on the whole sample. The biases for $\log g$ and [M/H] are relatively high; 

\item RED564 $\rightarrow$ similar to BLUE390. The biases were calculated on the whole sample and a high bias in $\log g$ was found;

\item RED860 $\rightarrow$ a linear relation was derived for the $\log g$ correction to the giant sub-sample as it showed a non-constant but linearly decreasing offset between the AMBRE and reference parameters.

\end{itemize}

While for the most part the parameter determination for each setup showed reasonable agreement with the Reference parameters, and in some cases, excellent agreement, there were also some large offsets and some large dispersions for some parameter sets. There were several contributing factors to this that did not necessarily apply to each set up. For instance:

\begin{itemize}

\item The BLUE setups are not as well represented by the synthetic spectra due to the greater number of spectral features, both atomic and molecular. Hence there were greater difficulties normalisation due to strong and highly depressed features resulting in relatively high systematic biases as above; 

\item In the RED the H$_{\alpha}$ and H$_{\beta}$ features were strong for some spectral types also creating difficulties in applying a generalised normalisation procedure. However overall the RED setups performed better than the BLUE; 

\item RED564 had a very small reference sample, particularly for the giants, with which to calibrate the pipeline resulting in calculating biases (very high bias in $\log g$) based on the whole sample; 

\item For RED860 despite selecting regions expected to be low in telluric contamination it was impossible to avoid this completely and so parameter determination is more uncertain.

\end{itemize}

\begin{figure*}[!ht]
\centering
\begin{minipage}{90mm}
\hspace{-0.5cm}
\includegraphics[width=100mm]{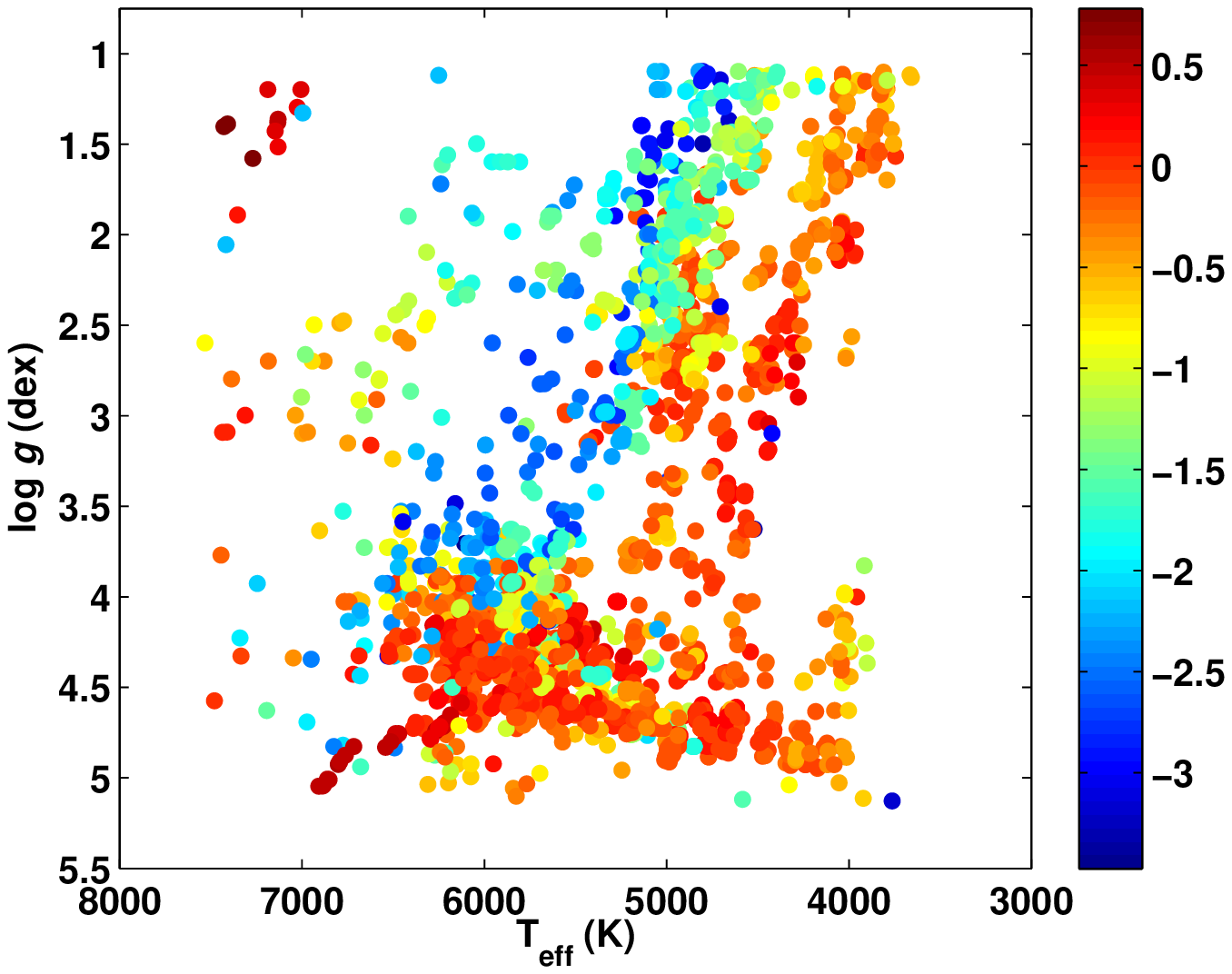}
\caption{HR Diagram with Metallicity colour map for the final cleaned sample for RED580.}\label{fig:clean_hrdm_red580}
\end{minipage}
\hfill
\begin{minipage}{90mm}
\hspace{-0.5cm}
\includegraphics[width=100mm]{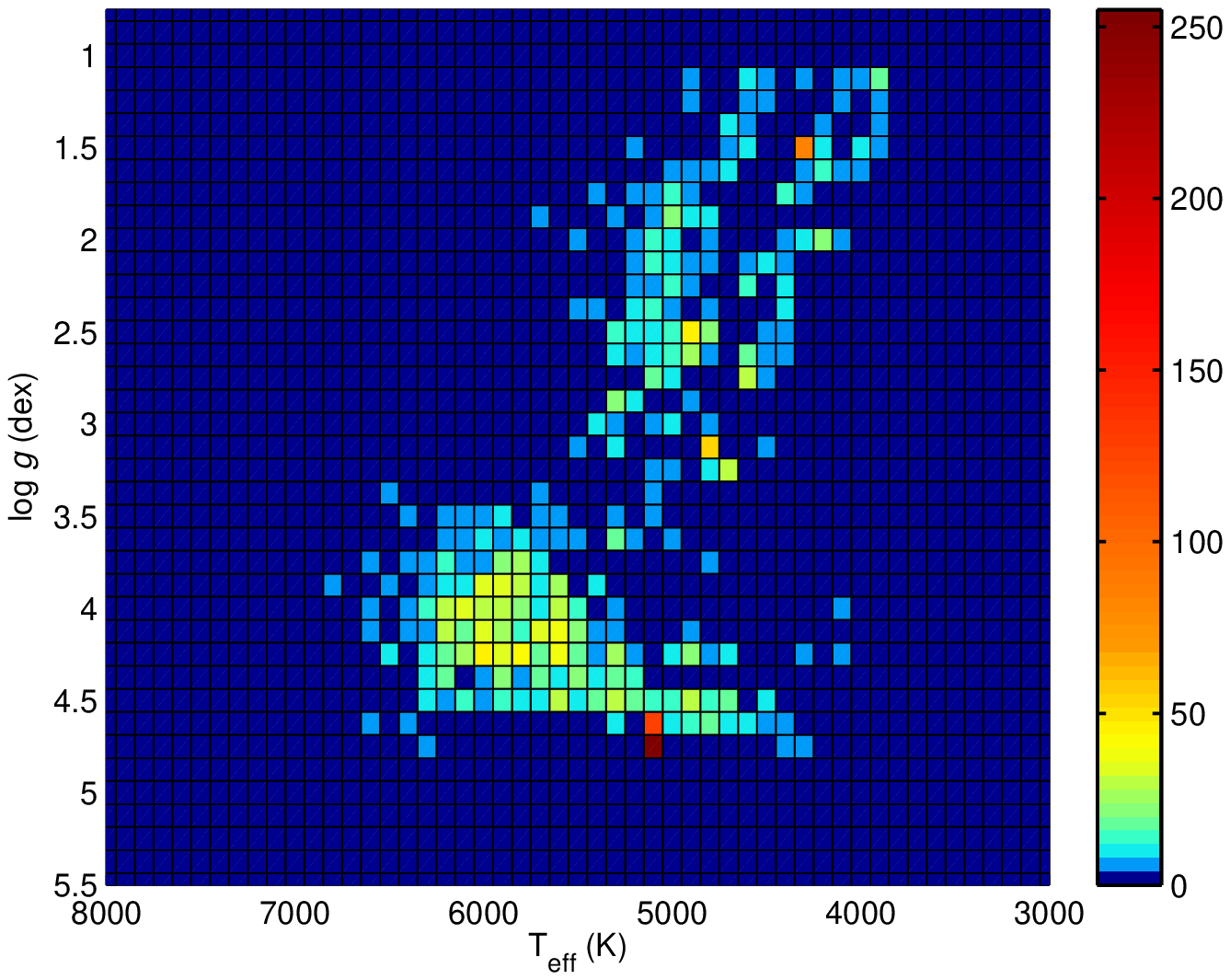}
\caption{Number Density HR Diagram of uncorrected cleaned RED580 sample showing empirical $\log g$ separation between dwarfs and giants.}\label{fig:clean_numden_red580}
\end{minipage}
\end{figure*}

As an example of the final outcome of the above process, the HR Diagram of the final cleaned per spectrum sample for RED580 is shown in Figure~\ref{fig:clean_hrdm_red580}. The metallicity is also provided as a colourmap. The most interesting feature is the split in the giant branch, which is clearly due to a division between metal-poor and metal-rich stars.

Figure~\ref{fig:clean_numden_red580} shows the RED580 HR Diagram as a number density plot. This more clearly represents the distribution of the RED580 sample showing the concentration of stars on the upper main sequence and at $\log g \sim 2.5$~dex on the Giant Branch. Also the separation between the Giant Branch and Main Sequence is distinct. This, and similar plots for the other five setups, were used to empirically define the Giant-Dwarf $\log g$  threshold.

\section{Internal \& External Errors}\label{sec:intexterrors}
The standard deviations listed in Table~\ref{tab:biases} have been adopted as the external error associated with each sub-sample and reported as such for the final submission to ESO. See also Table~\ref{tab:esotab_descrip} presenting the columns delivered to ESO. Defined by sub-sample, the standard deviations reflect the inherent difficulties in parameter determination for these setups and stellar types. Here, as for the FEROS and HARPS analyses, the standard deviation of the difference between the AMBRE and reference values is defined as the external error, where we mean a comparison to an external reference source as opposed to the more strict statistical definition of the uncertainty in repeated measurements of the same object \citep{Drosg2009}. Similarly, by internal error we mean uncertainties inherent in our method (S/N, continuum placement) and include for our purposes the spread in the repeated measurements of a single object. This is broader than the strictly statistical definition of the internal uncertainties which are only those inherent in a single measurement \citep{Drosg2009}.

We carried out an exploration of the internal errors present in our method. Drawn from the AMBRE:FEROS analysis, the variation in parameters associated with the iterative change in normalisation between the 9th and 6th iterations was shown for UVES to be consistently neglible ($\Delta T_{\textrm{eff}}\sim10$~K, $\Delta\log g\sim0.004$~dex, $\Delta$[M/H]$\sim0.003$~dex, $\Delta$[$\alpha$/Fe]$\sim0.002$~dex) across the range of S/N. 

The per parameter error provided by MATISSE is estimated from the S/N. However a better representation of the internal consistency of MATISSE with S/N is given by the analysis of the repeatability of parameters from multiple spectra of the same star across S/N. This assesses how well independent measurements of the same object agree when they are analysed by this automated process and hence treated in a consistent manner. This analysis is summarised below and the resulting relations are used to define the internal error for the UVES sample delivered to ESO as reported in Table~\ref{tab:esotab_descrip}.

\begin{figure}[!ht]
\centering
\begin{minipage}{90mm}
%\vspace{-0.5cm}
\hspace{-0.5cm}
\includegraphics[width=100mm]{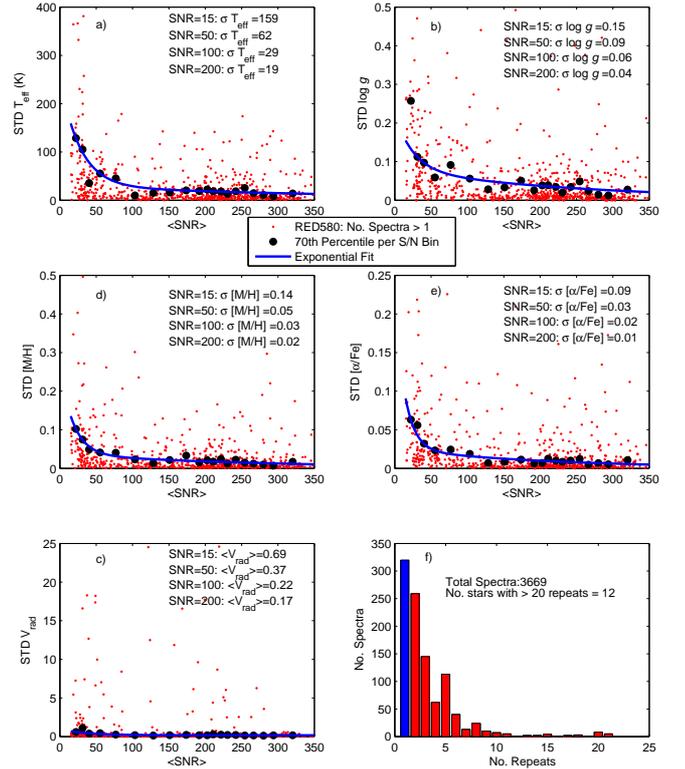}
\vspace{-1.0cm}
\caption{Analysis of repeated measurements for RED580 showing the standard deviation of parameters with S/N for a) $T_{\textrm{eff}}$; b) $\log g$; c) Vrad; d) [M/H] and e) [$\alpha$/Fe]. The histogram of number of spectra per number of repeats is shown in f). Individual stars are shown as red dots. The 70th percentile of the S/N bins are shown as black dots. The exponential fit to the black dots are shown as a blue line.}\label{fig:red580_repeats}
\end{minipage}
\end{figure}

Figure~\ref{fig:red580_repeats} shows this exploration of the RED580 repeats sample per parameter as a function of S/N. The repeats sample was found by identifying each star with multiple associated spectra using a radius cone search of 1.8'' about the coordinates of the first instance of the star in the spectra list. This was carried out using the final cleaned sample of spectra. 

The mean S/N and the standard deviation in each parameter for each star was then obtained. These are shown as red points in Figure~\ref{fig:red580_repeats}. In S/N bins of $\sim$15, the 70th percentile of the standard deviations were determined, shown as black dots. A clear increase to lower S/N is shown being approximately exponential for S/N$<$30. An exponential fit to these black points was made to derive the final relation (shown in blue).

For S/N greater than $\sim$200 the relation is approximately constant, and certainly above 350 the bins begin to suffer from small number statistics. Therefore an upper limit in S/N is imposed above which the internal error is taken as the value calculated at that upper limit. The upper limit selected per setup based on the respective diagnostic diagrams is given in Column 7 of Table~\ref{tab:repeatscoeff}.

\begin{table*}[htbp]
\tabcolsep=0.10cm
\caption{Per setup and per parameter internal error relation coefficients.}
\begin{tabular}{lccccc|ccccc|ccccc}
\hline\hline
 & & \multicolumn{ 3}{c}{BLUE346 } & & & \multicolumn{ 3}{c}{BLUE390} & & & \multicolumn{ 3}{c}{BLUE437 } & \\ 
 &  & $T_{\textrm{eff}}$ & $\log g$   & $[$M$/$H$]$   & $[\alpha/$Fe$]$   &  & $T_{\textrm{eff}}$ & $\log g$   & $[$M$/$H$]$   & $[\alpha/$Fe$]$   &  & $T_{\textrm{eff}}$ & $\log g$   & $[$M$/$H$]$   & $[\alpha/$Fe$]$   \\ 
\hline
c$_1$  &  & 95.0912 & 0.4931 & 0.3226 & 0.6522 &  & 0.0000 & 0.3188 & 0.2664 & 0.0790 &  & 54.9590 & 4.6684 & 0.0944 & 0.6067 \\ 
c$_2$  &  & -0.0539 & -0.1057 & -0.0935 & -0.1263 &  & -1.5110 & -0.0381 & -0.0358 & -0.0814 &  & -0.0295 & -0.2533 & -0.0289 & -0.1572 \\ 
c$_3$  &  & 0.5179 & 0.0057 & 0.0020 & 0.0041 &  & 219.4735 & 0.0022 & 0.0001 & 0.0754 &  & 0.0007 & 0.1447 & 0.0000 & 0.0234 \\ 
c$_4$  &  & 0.0133 & -0.0001 & 0.0056 & -0.0065 &  & -0.0325 & 0.0092 & 0.0316 & -0.0250 &  & 0.0576 & -0.0251 & 0.0758 & -0.0183 \\ 
 & S/N\tablefootmark{a} &  &  &  &  & S/N &  &  &  &  & S/N &  &  &  &  \\ 
$\sigma_{(S/N_{L})}$  & 15 & 43.0 & 0.127 & 0.082 & 0.102 & 15 & 134.7 & 0.189 & 0.156 & 0.075 & 15 & 35.3 & 0.204 & 0.061 & 0.075 \\ 
$\sigma_{(S/N_{U})}$ & 125 & 2.8 & 0.006 & 0.004 & 0.002 & 125 & 3.8 & 0.011 & 0.007 & 0.003 & 125 & 2.2 & 0.006 & 0.003 & 0.002 \\ 
\hline
&  &  &  &  &  &  &  &  &  &  &  &  &  &  &  \\ 

 & & \multicolumn{ 3}{c}{RED564 } & & & \multicolumn{ 3}{c}{RED580 } & & & \multicolumn{ 3}{c}{RED860 } & \\ 
 &  & $T_{\textrm{eff}}$ & $\log g$   & $[$M$/$H$]$   & $[\alpha/$Fe$]$   &  & $T_{\textrm{eff}}$ & $\log g$   & $[$M$/$H$]$   & $[\alpha/$Fe$]$   &  & $T_{\textrm{eff}}$ & $\log g$   & $[$M$/$H$]$   & $[\alpha/$Fe$]$   \\ 
\hline
c$_1$  &  & 0.0000 & 0.0000 & 7410.0224 & 0.0021 &  & 227.0935 & 0.1456 & 0.2299 & 0.1971 &  & 0.0000 & 0.0000 & 0.0000 & 0.0000 \\ 
c$_2$  &  & -7.1585 & -9.7333 & -0.0138 & -0.0011 &  & -0.037 & -0.0401 & -0.0553 & -0.0719 &  & -3.5645 & -0.3207 & -0.3583 & -0.3416 \\ 
c$_3$  &  & 282.8015 & 0.1872 & -7409.7631 & 0.1988 &  & 29.6691 & 0.0793 & 0.0372 & 0.0254 &  & 182.9632 & 0.3093 & 0.4260 & 0.0928 \\ 
c$_4$  &  & -0.0325 & -0.0157 & -0.0138 & -0.0368 &  & -0.0023 & -0.0038 & -0.0036 & -0.0047 &  & -0.0070 & -0.0060 & -0.0090 & -0.0066 \\ 
 & S/N &  &  &  &  & S/N &  &  &  &  & S/N &  &  &  &  \\ 
$\sigma_{(S/N_{L})}$  & 20 & 147.6 & 0.137 & 0.197 & 0.097 & 15 & 159.1 & 0.138 & 0.136 & 0.091 & 15 & 164.6 & 0.283 & 0.372 & 0.084 \\ 
$\sigma_{(S/N_{U})}$ & 175 & 1.0 & 0.012 & 0.023 & 0.002 & 200 & 18.7 & 0.037 & 0.018 & 0.010 & 200 & 63.6 & 0.125 & 0.110 & 0.035 \\ 
\hline
\end{tabular}
\tablefoot{
\tablefoottext{a}{S/N value used for each parameter per setup.}
}
\label{tab:repeatscoeff}
\end{table*}

Table~\ref{tab:repeatscoeff} sets out the coefficients to the exponential fits for each UVES setup for each parameter, as well as the S/N upper limit and corresponding constant value. The value at the lower S/N limit imposed for the cleaning of each sample is also given. The equation of the exponential fit is:  $\sigma_{\theta} = {\bf c_1}exp({\bf c_2}S/N) + {\bf c_3}exp({\bf c_4}S/N)$.

Comparison of the external and internal errors can be made by considering the bias columns of Table~\ref{tab:biases} ($\sigma_{parameter}$) with the rows of internal error estimates at the low S/N limit ($\sigma_{(S/N_{L})}$) of Table~\ref{tab:repeatscoeff}. The lower S/N limit, S/N$>$15 (S/N$>$20 for RED564), of the internal errors generally approach the external error constant values. Thus while the external errors may seem overestimated, they agree reasonably well with the internal errors at low S/N, for which a significant number of the reported UVES sample lies as shown in Figure~\ref{fig:setup_hists}.

The internal and external errors are deliverables for ESO in the respective columns, ERR\_INT\_ and ERR\_EXT\_, per parameter as listed in Table~\ref{tab:esotab_descrip}. The cleaned, bias-corrected dataset with these errors comprise the final parameters delivered to ESO and are used in the following discussion.

\subsection{Contamination of Final Sample}
The AMBRE analysis relies on a series of tests on the radial velocity, spectral FWHM and parameter measurements in terms of errors and goodness of fit to reject spectra from the sample. No specific test is carried out to identify particular types of objects. Binary systems in particular are not searched for specifically, however it is expected that for some spectroscopic binaries, the multi-component spectrum would be a poor fit to the synthetic spectrum and be rejected by a high $\log(\chi^2)$.

The radial velocity CCF is likely to be broad or double-peaked (multi-peaked) for a binary (multiple) system, activating also larger errors and poorer assesment of the CCF fits. While not used here, detecting such objects by tests on the CCF is a tool that could be developed and is planned for the future.

Variable objects could also be detected if multiple observations are available within the sample over some pertinent timeframe but again this is not explored here, as the analysis is per spectrum, not per star.

Certainly the coordinates and object names provided within the spectral headers can be used to search within variable star catalogues and thus detect any known variables within the sample. However as the parameters provide a snapshot of the star at that time they need not necessarily be rejected.

Simple searches on the final accepted sample looking for key naming nomenclature yields some idea of residual contamination by non-stellar or variable objects. For instance, searching for `V' in object name yields some 85 spectra with object name of the form `VXXX Constellation' (e.g. V580\_Cen) as per the General Catalogue of Variables.

Table~\ref{tab:contaminants} provides a list of non-stellar objects which have been found by target name within the AMBRE:UVES sample. It must be noted that the motivation for naming an observation a certain way can cover observing candidate objects, reference objects, correcting pointing errors and so forth. Providing a non-stellar object classification as an object name is no guarantee that it is in fact non-stellar and parameters attributed to something named as `SN' may be a valid parameter set for what is actually a stellar object. 

\begin{table}[htbp]
\centering
\caption{Spectral, V$_{rad}$ and $T_{\textrm{eff}}$ count of non-stellar objects within AMBRE:UVES}
\begin{tabular}{lccc}
\hline\hline
Non-Stellar &    No. Spectra   &   No. V$_{rad}$   &   No. $T_{\textrm{eff}}$ \\ 
\hline
SN  & 405 & 220 & 20 \\ 
GRB & 146 & 101 & 2 \\ 
QSO  & 149 & 94 & 2 \\ 
GRDG8  & 35 & 28 & 0 \\ 
Nova  & 114 & 84 & 2 \\ 
CV  & 27 & 25 & 1 \\ 
\hline
\textbf{All UVES} & \textbf{51921} & \textbf{36881} & \textbf{12403} \\ 
Non-Stellar & 876 & 552 & 27 \\ 
Percentage(\%) & 1.7 & 1.5 & 0.2 \\ 
\hline
\end{tabular}
\label{tab:contaminants}
\end{table}

\begin{figure*}[!ht]
\centering
\begin{minipage}{180mm}
\hspace{-2cm}
\vspace{-1cm}
\includegraphics[width=215mm]{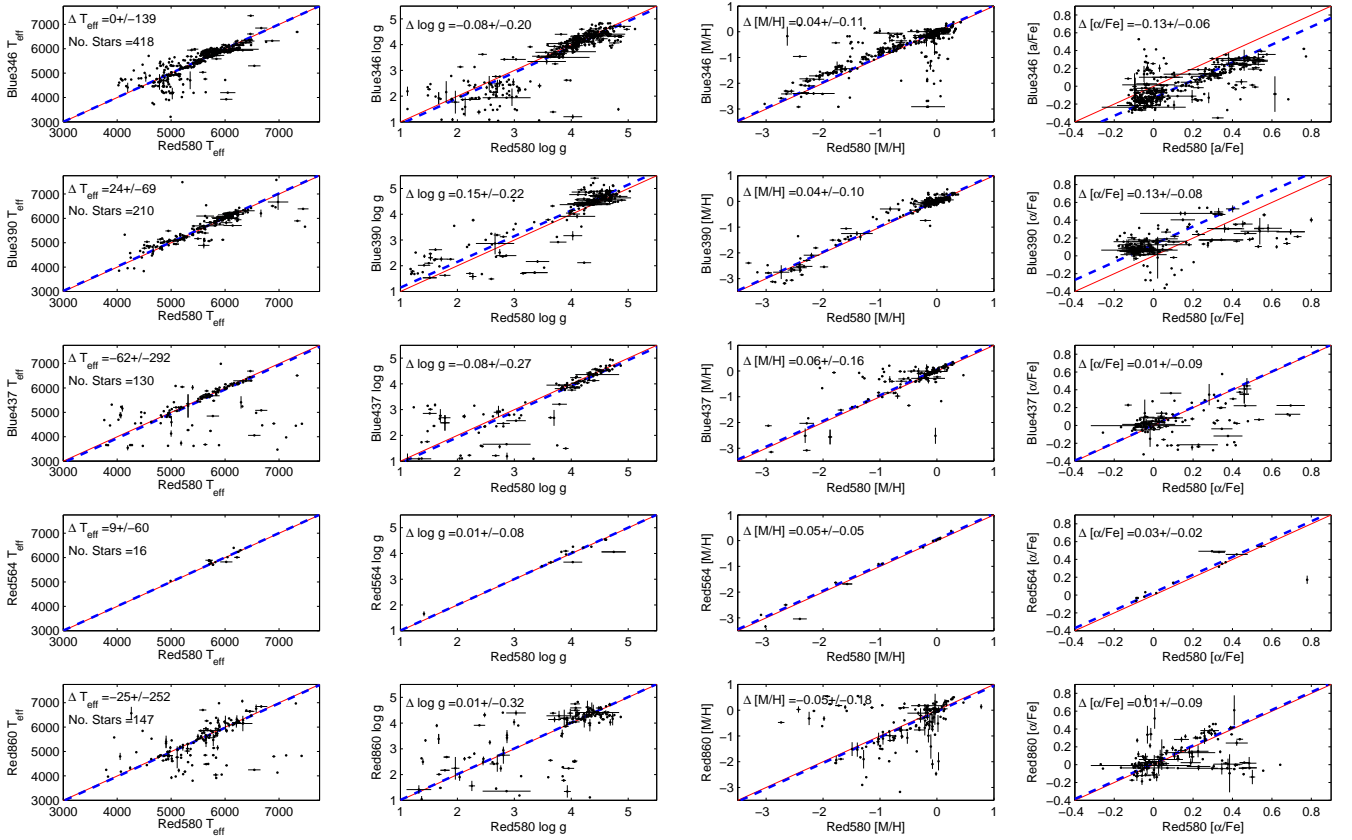}
\caption{RED580 parameters versus the other five setup parameters for each cross-match sample with the number of stars, bias and dispersion listed. The red line is the 1:1, while the blue dashed line shows the bias.}\label{fig:intersetupcomp}
\end{minipage}
\end{figure*}

As ESO provided all observed spectra within the specificed timeframe, we return the same without removing form the list any spectra we have not classified, including non-stellar objects. As such we rely on the various quality criteria for catching such objects and rejecting any parameters we have determined for them. As seen in Table~\ref{tab:contaminants}, we do return some parameters for potential non-stellar objects. For example, for objects named as 'SN' or  `SN-YEARLetter' (e.g. SN-1987A), indicating they are Supernova candidate observations, we report parameters for 20 out of 405. Similar for 2 out of 149 GRBs (Gamma Ray Burst objects), 2 out of 149 QSOs (Quasi-Stellar Objects), 2 out of 114 nova and 2 out of 27 Cataclysmic Variables (CV). In total this comes to 0.2\% of the sample of AMBRE:UVES that have reported stellar parameters. Visual inspection of each of these spectra showed that these objects typically were of low S/N (S/N$\sim$25) with few spectral features. However they did not activate any of the rejection criteria and so the parameters were reported. Such a small percentage of contaminants is expected given the blind nature of the analysis and the necessity to define empirical limits for the rejection criteria for which some non-stellar objects are able to pass.

A V$_{rad}$ is reported for a much larger sample of these non-stellar objects (1.5\%) such as objects within the irregular dwarf galaxy, GRDG8. For both the reported parameters and V$_{rad}$, the spectra of these objects have not failed the quality criteria and as said above a target name, while indicative, is not necessarily definitive as to the type of object. From this investigation the potential contamination of non-stellar objects within the stellar parameters that we have reported is less than a percent.

%Thus a simple object name search finds a handful of possible contaminants comprising less than 3\% of the accepted sample.

\section{Inter-setup Comparison}\label{sec:intersetupcomp}
The analysis undertaken thus far treats each setup as an individual dataset, for which approximately the same process was followed. A test of the robustness of the AMBRE analysis is the comparison of the results between the setupS.

For this exploration stars in common between RED580 (the largest sample) and the other five setups were compared. The cross-match on the spectra was carried out using a very restricted coordinate radius of 0.18'' for as clean a sample as possible. The cross-match between RED580 and RED564 required a less restricted radius of 1.8'' as the intersecting sample is very small. It is likely that some non-trivial fraction of the cross-match samples were observed simultaneously between the RED and BLUE particularly, and so the S/N are likely to differ between the setupS depending on the particular programme goals.

Figure~\ref{fig:intersetupcomp} shows the 1:1 diagrams per parameter for this comparison. For each star multiple spectra were possible within each setup, hence the error bars are the spread in parameters for the spectra in that setup, or the external error if only one spectrum was found. The number of stars for each cross-match, and the bias and dispersion per parameter for the cross-matches are given in the respective panels. The bias and dispersion are calculated based on a 1~$\sigma$ clip of the sample. The red line shows the 1:1 line, while the blue dashed line is the bias.
\begin{figure*}[!ht]
\centering
\begin{minipage}{180mm}
\hspace{-2.5cm}
\includegraphics[width=240mm]{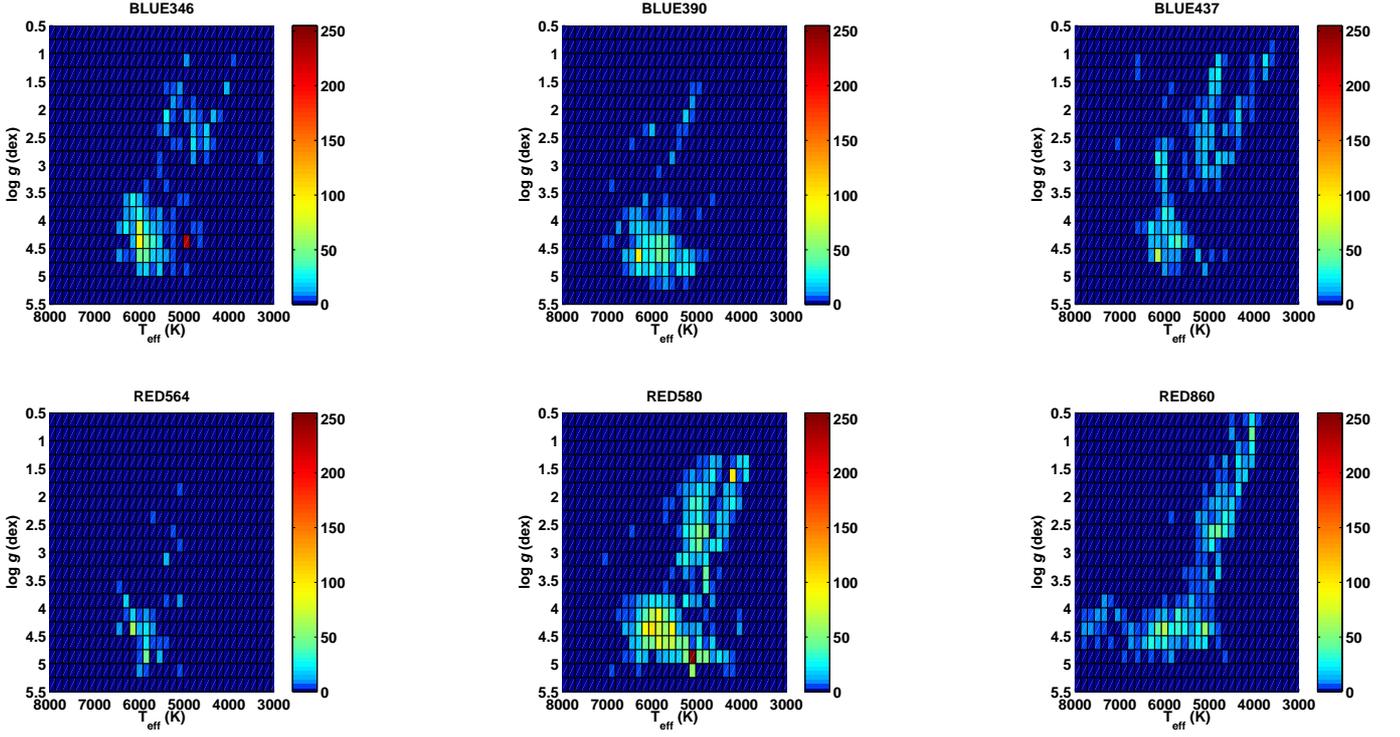}
\caption{Hess diagrams for the cleaned corrected samples for the six setups. The number density colourbar for each is set to the same scale.}\label{fig:numdens_all}
\end{minipage}
\end{figure*}

\begin{figure*}[!ht]
\centering
\begin{minipage}{180mm}
\hspace{-2.5cm}
\includegraphics[width=240mm]{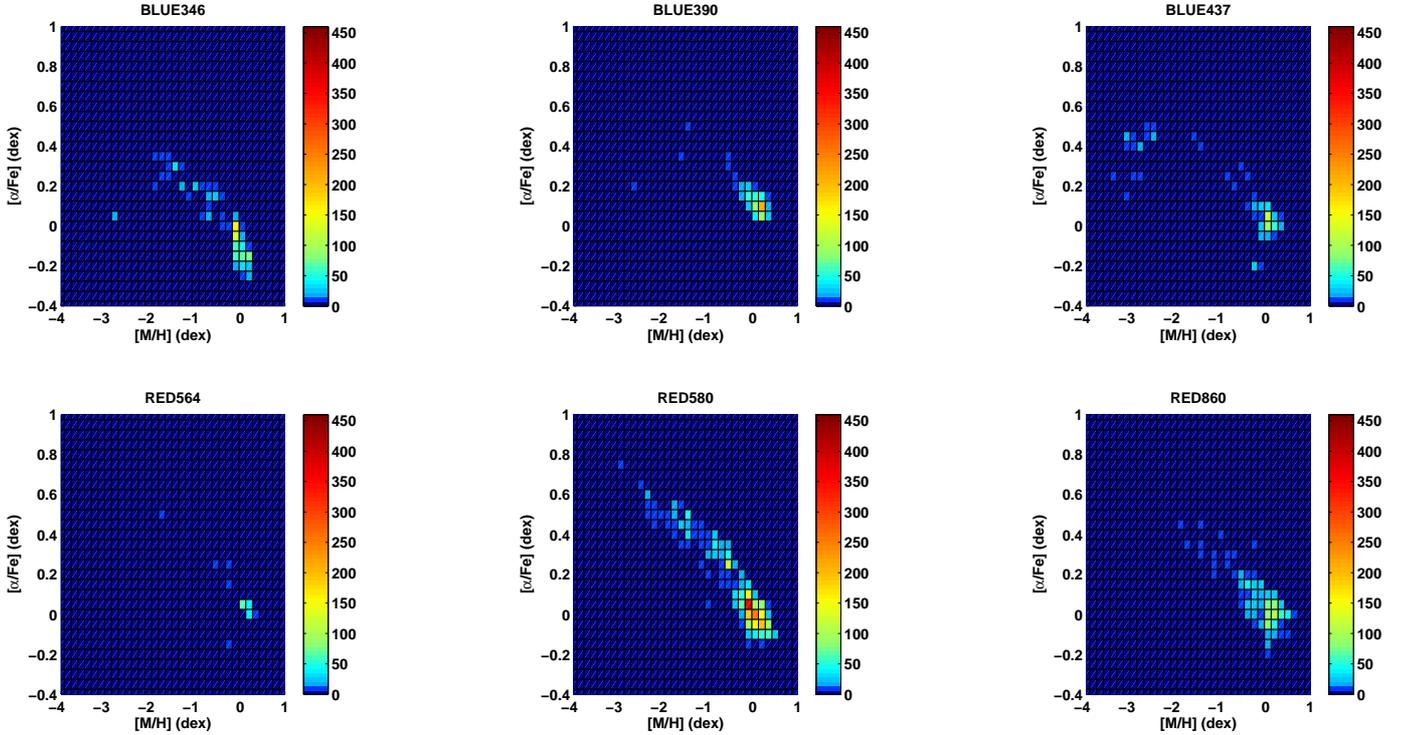}
\caption{[M/H] versus [$\alpha$/Fe] as number density plots for the six setups as listed for the cleaned corrected samples. The number density colourbar for each is set to the same scale.}\label{fig:numdens_ma_all}
\end{minipage}
\end{figure*}

Visual inspection shows a relatively good agreement in parameters between the setups with clear trends along the 1:1 relations, and associated biases being generally very small. Dwarfs are in better agreement than giants generally, with the giant samples showing greater dispersion. The relations in [$\alpha$/Fe] show high scatter although the 1~$\sigma$ dispersions are quite reasonable. Both $T_{\textrm{eff}}$ and [M/H] are overall in very good agreement between the setups.

There are some clear systematic mis-classifications, particularly between RED580 and BLUE437 seen as a horizontal line of stars with RED580 $T_{\textrm{eff}}$ greater than 5000~K having a corresponding BLUE437 $T_{\textrm{eff}}$ (RED860 $T_{\textrm{eff}}$) less than 5000~K. A similar feature (though not necessarily the same stars) is seen in the RED580 and RED860 comparison.

The largest relative bias is between RED580 and BLUE390 in $\log g$ at a value of -0.15~dex although this is within the dispersion of 0.22~dex. BLUE390 did not have a well distributed calibration sample (see Section~\ref{sec:calibsamps}) lacking giants in particular. Thus while the dwarfs in common with RED580 look in reasonable agreement, the significantly smaller sample of giants in common show some offset and are quite dispersed.

Even though the sample between RED580 and RED564 is very small there is excellent agreement between the parameters in the cross-matched sample.

Based on these comparison no inter-setup corrections have been applied as each setup was calibrated by the same method, albeit not the exact same sample of key stars. The good agreement shown here is very satisfactory and strengthens the present parameterisation.

\section{Discussion}\label{sec:discussion}

Figures~\ref{fig:numdens_all} and \ref{fig:numdens_ma_all} show the HR diagram and Metallicity versus [$\alpha$/Fe] abundance ratio as number density plots for the final cleaned, corrected samples for each UVES setup. They illustrate the quite different samples of stars recovered for each setup. RED580 gives the best coverage of both the main sequence and giant branches. Both BLUE437 and RED860 include both main sequence and giant stars, but quite different morphology for the upper main sequence and metal-poor giant branches in each case. Both BLUE390 and RED564 seem to mainly comprise solar-metallicity dwarf stars, with a greater number recovered for BLUE390.

The relation of [$\alpha$/Fe] to [M/H] is well produced in all setups, if lacking a continuum to the metal-poor in most cases. However the RED580 shows a very clear well-defined relation.

\begin{figure}[!ht]
\centering
\begin{minipage}{90mm}
\hspace{-0.5cm}
\includegraphics[width=95mm]{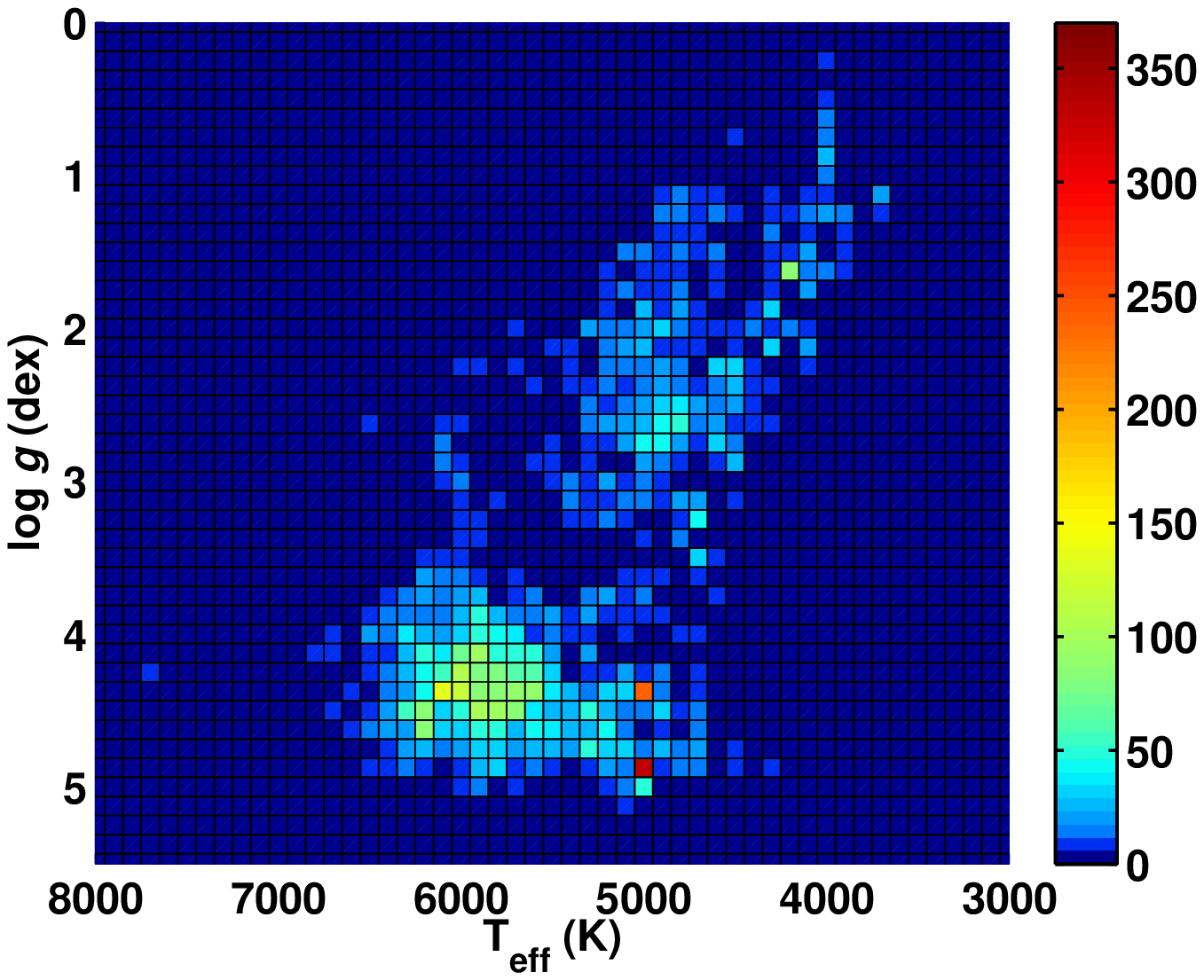}
\caption{Hess diagram of the final combined, cleaned and bias-corrected UVES dataset.}\label{fig:numdens_all_cor}
\end{minipage}
%\hfill

\begin{minipage}{90mm}
\hspace{-0.5cm}
\includegraphics[width=95mm]{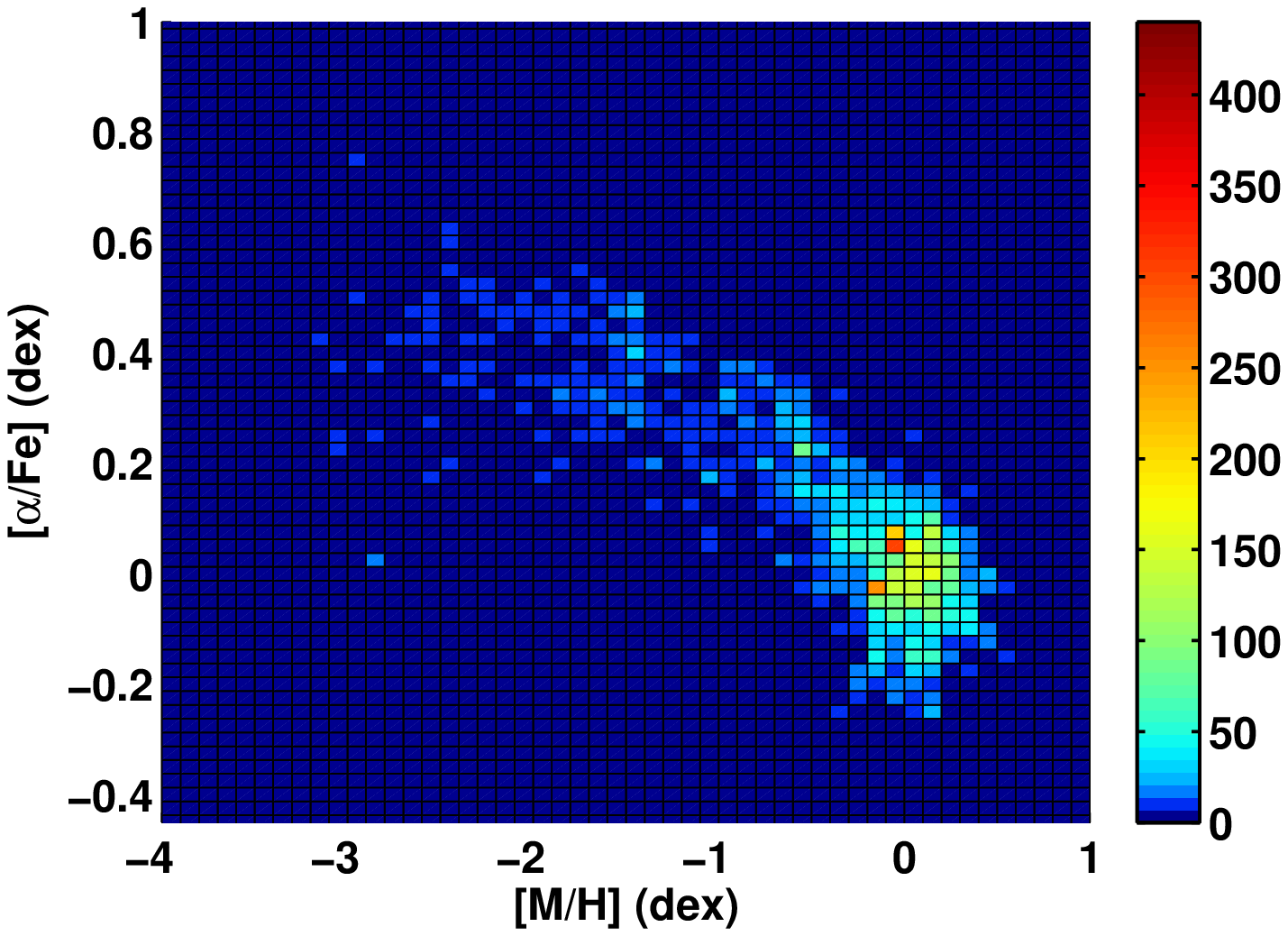}
\caption{As for Figure~\ref{fig:numdens_all_cor} but showing the Metallicity versus [$\alpha$/Fe].}\label{fig:numdens_all_mhafe_cor}
\end{minipage}
\end{figure}

Figure~\ref{fig:numdens_all_cor} shows the HR Diagram as a number density plot (Hess Diagram) for the cleaned combined UVES sample with the bias corrections applied. This figure can be viewed as an ``addition'' of the 6 plots shown in Figure~\ref{fig:numdens_all}. It can be seen that, with the parameters scaled to the calibration stars, the corrected sample provides a coherent consistent sample with the separation between dwarfs and giants clearly defined. 

Figure~\ref{fig:numdens_all_mhafe_cor} shows the metallicity versus [$\alpha$/Fe] distribution for the final sample also as a number density plot. This relation is very well-defined showing the expected [$\alpha$/Fe] enrichment at low metallicity to solar values at solar metallicity. 

\begin{figure}[!hp]
\begin{minipage}{90mm}
\centering
%\hspace{-1.3cm}
\includegraphics[width=65mm]{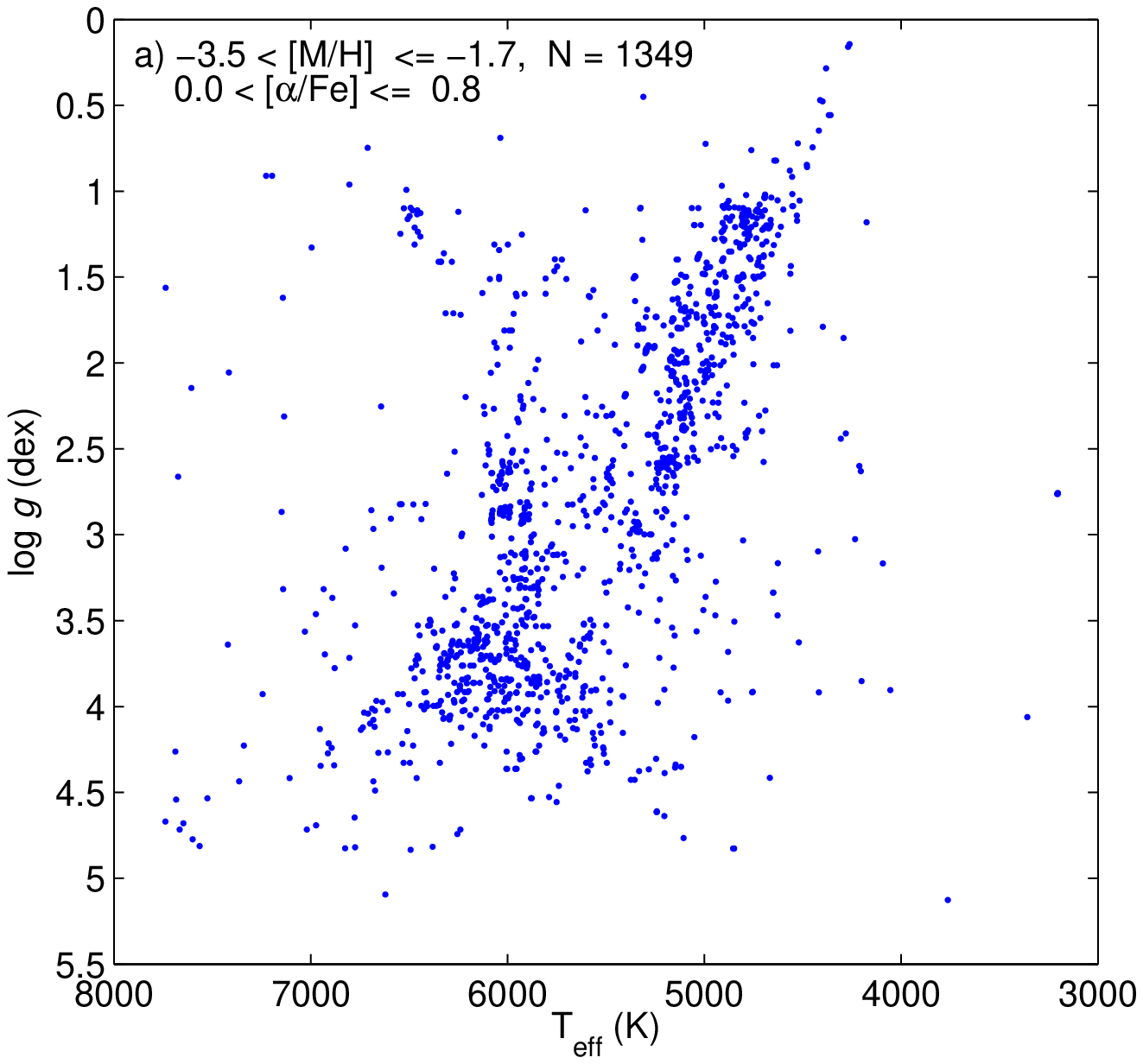}
\vspace{-0.2cm}
\end{minipage}
%\hfill
\begin{minipage}{90mm}
\centering
%\hspace{-1.3cm}
\includegraphics[width=65mm]{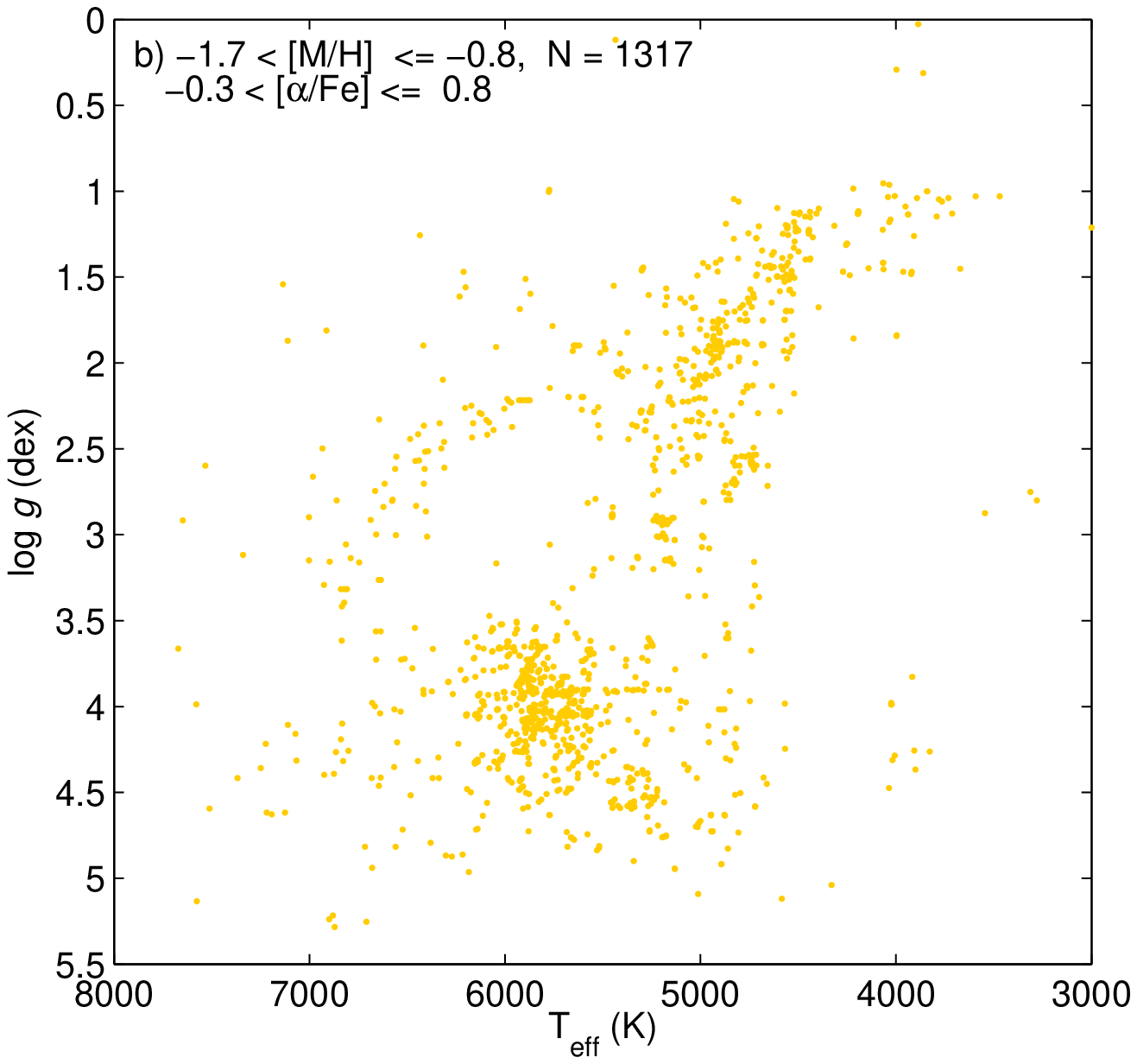}
\vspace{-0.2cm}
\end{minipage}
%\hfill
\begin{minipage}{90mm}
\centering
%\hspace{-1.3cm}
\includegraphics[width=65mm]{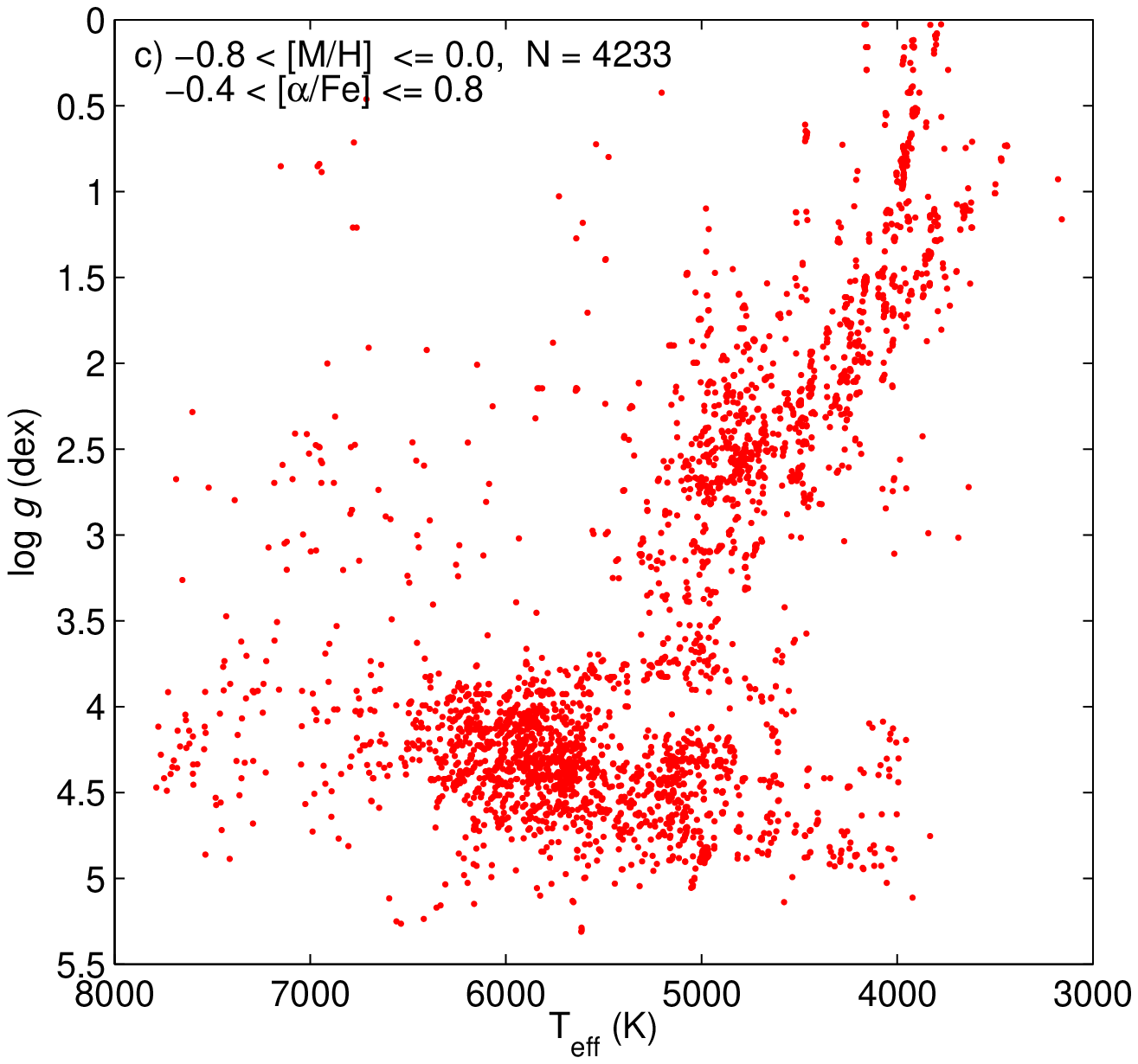}
\vspace{-0.2cm}
\end{minipage}
%\hfill
\begin{minipage}{90mm}
\centering
%\hspace{-1.3cm}
\includegraphics[width=65mm]{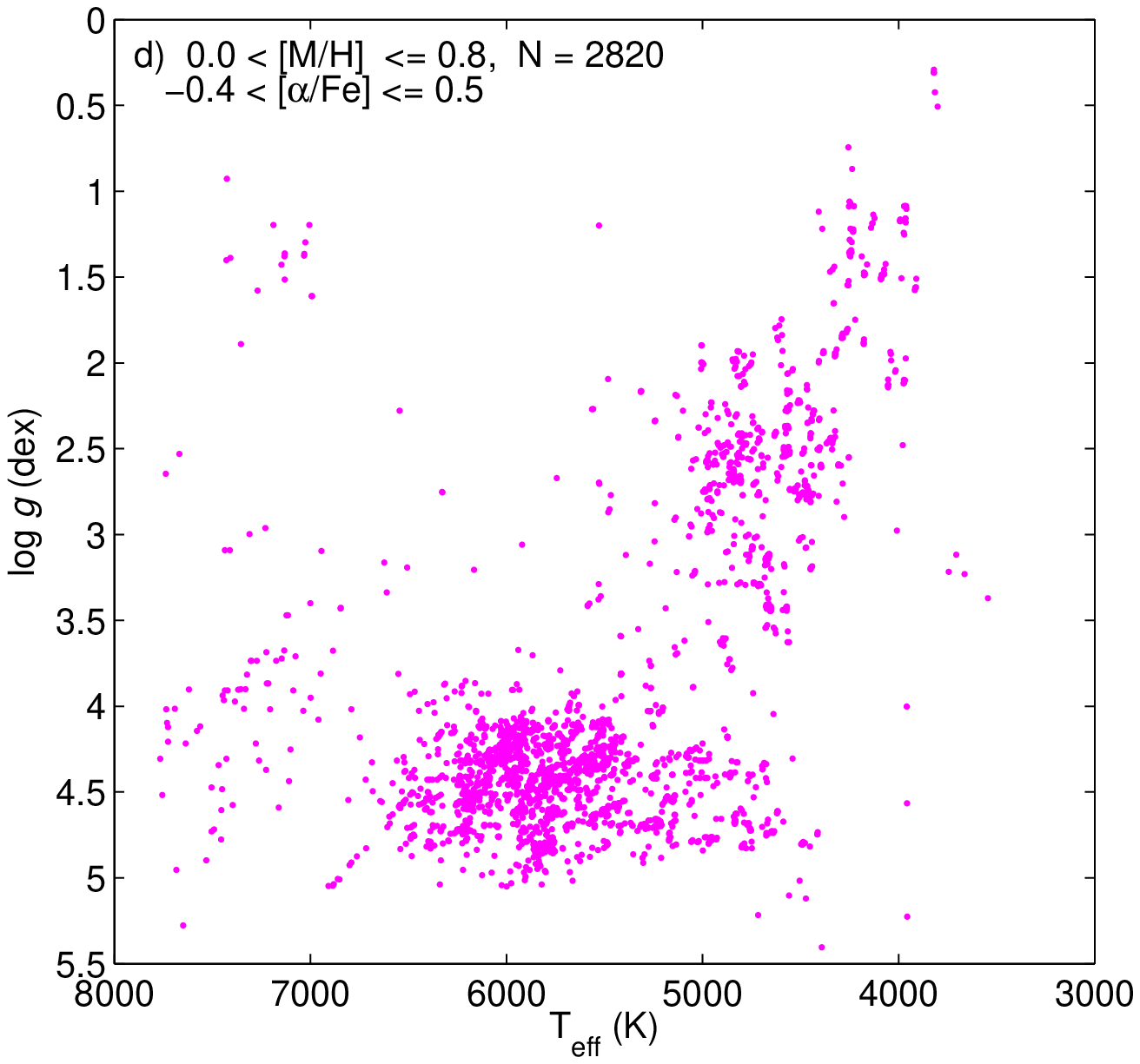}
\vspace{-0.2cm}
\end{minipage}
\caption{HR Diagrams of the final cleaned and corrected UVES dataset binned in metallicity as specified with associated ranges in [$\alpha$/Fe] such that: a) the very metal-poor bin, b) the metal-poor bin, c) the sub-solar bin, and d) the metal-rich bin.}\label{fig:hrd_metbin_all}
%\end{minipage}
%\end{minipage}
\end{figure}

A distinct feature in the final dataset is the rather inflated Giant Branch in Figure~\ref{fig:numdens_all_cor} which also shows a clear bifurcation in the upper section. This in fact corresponds to the range of metallicities present in the dataset and is illustrated in Figure~\ref{fig:hrd_metbin_all}, which shows the final UVES sample in a series of HR Diagrams binned in metallicity as specified in each diagram.

%\begin{minipage}{90mm}

The main effect see in Figure~\ref{fig:hrd_metbin_all} between the metallicity bins is the movement of the Giant Branch to cooler temperatures with increasing metallicity as is expected. The very metal-poor bin (a) shows the largest scatter and least well defined branches of stellar evolution. With increasing metallicity the branches become more clearly defined particularly for the bin shown in (c). This movement in the Giant Branch constructs the inflated, bifurcated morphology seen in Figure~\ref{fig:numdens_all_cor} comprising a peak in solar metallicity with a second less populated, more scattered metal-poor peak.

The range of [$\alpha$/Fe] values for each metallicity bin is also specified in each panel. While each metallicity bin has a wide range in [$\alpha$/Fe], there is a global trend from enhanced [$\alpha$/Fe] to depleted [$\alpha$/Fe] with increasing metallicity, also as expected and as seen in the complete dataset in Figure~\ref{fig:numdens_all_mhafe_cor}.

This series of figures shows that the construction of the final UVES dataset from the spectra of the six setups has produced a single, coherent, consistent sample that can be deconstructed by metallicity and [$\alpha$/Fe] in agreement with the expected relations within stellar populations. %This is a clear validation of the robust parameterisation and calibration processes undertaken within the AMBRE:UVES analysis.

As illustrated in the above figure, as a high resolution instrument on a 10~m class telescope, UVES has been used to extend the forefront of stellar populations research. The detection of the first stars is ideally suited to this instrument, and the metal-poor sample of observations is clearly evident in Figure~\ref{fig:hrd_metbin_all}a.

\begin{figure}[!ht]
\centering
\begin{minipage}{90mm}
\hspace{-0.9cm}
\includegraphics[width=105mm]{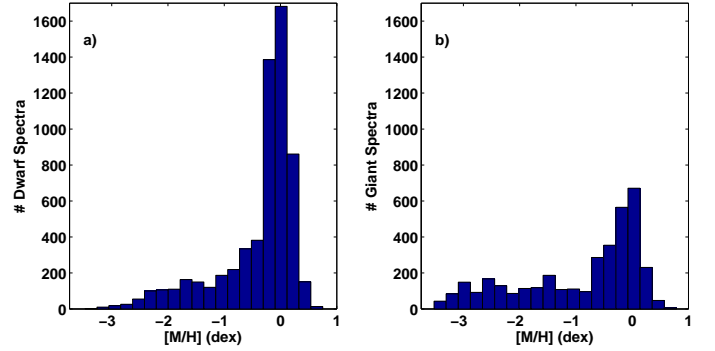}
\caption{Metallicity Distribution of AMBRE:UVES parameterisation for a) Dwarfs ($\log g \ge 3.5$), and b) Giants ($\log g < 3.5$).}\label{fig:hist_all_mh}
\end{minipage}
\end{figure}

However the sample is not evenly distributed between these metallicity bins. Figure~\ref{fig:hist_all_mh} instead shows the metallicity distribution for the entire UVES sample separated by dwarfs ($\log g \ge 3.5$) and giants ($\log g < 3.5$) respectively. The sample is dominated by solar metallicity stars, reflected in both the dwarf and giant subsamples. However there are non-neglible metal-poor tails to each distribution, with a secondary peak in both dwarfs and giants at [M/H]$\sim$-1.5~dex. There is a potential third peak in the giant sample at -2.5~dex, possibly reflecting a bias towards searching for the most metal-poor stars in the more luminous giant population. From the sample of AMBRE:UVES spectra with [M/H] less than -2.0 there are 215 distinct ESO Observing Programme IDs most likely including the dedicated metal-poor programmes. Exploration of the AMBRE:UVES sample with comparison to these programmes will be investigated in the science follow up to the AMBRE parameterisation.

\section{Conclusion}\label{sec:conclusion}
The AMBRE:UVES parameterisation of the UVES archived spectra encompassed a challenging sample of spectra covering a diverse range of wavelengths and stellar types. The success of this analysis is that it has recovered a comprehensive dataset across the FGK parameter range consistent with expected stellar populations.

Each standard setup analysed shows the particular selection of stellar types for which that setup has been used by observers. RED580 is the most comprehensive in terms of the FGK HR Diagram. Others show a clear bias towards main sequence stars, for those stars for which parameters could be recovered. An example of the inherent observational biases within the UVES sample is the distinct metal-poor sample particularly within the giant sample.

From a total of 51921 spectra (78406 spectra if RED L and U spectra are considered separately), AMBRE:UVES reported parameters for 23.9\% of the sample, rejecting 76.1\%. The majority of rejections (38.3\%) were spectra with too broad FWHM$_{CCF}$ indicating that, if they are stars they are too hot or fast rotating for analysis by the parameter space and resolution configuration of the grid. 

The UVES sample as provided by ESO contained a veritable sm\"{o}rg\aa{}sbord of celestial objects observed over a 10 year period. From merely a visual scan of object names we found extra-galactic observations of gamma ray bursts and supernova, galactic observations of nova, nebula and variable stars, and solar system observations of planets and satellites. There was no prior sorting of the sample by ESO to provide only stellar objects to OCA. Therefore we relied on extensive automated tests on spectral quality and fitting (V$_{rad}$ errors and $\log \chi^2$) to provide a quantitative way to discard non-stellar spectra, as well as spectra with multiple stellar components. The sample reduced to our targetted FGK stars is a minority of the total UVES sample, and there is some small contamination by non-stellar objects in the final parameterised sample due to the automated and blind nature of our analysis.

The rejection process occurred in two stages, the first being objects which were assessed as not being capable of analysis, either due to quality issues (low S/N, technical issues) or being intrinsically outside the parameter space of the spectra grid (e.g. too broad spectral features). 32,306 (62.2\%) spectra were rejected at this first stage. While this is a high percentage to reject, the rejections comprise a great many non-stellar objects, as well as stellar objects for which our analysis is not designed, and reflects the great variety of objects observed by UVES. The second stage of rejections comprises stars for which parameterisation was potentially possible but quality criteria and error analysis implied that the results were not reliable. 7,212 spectra (13.9\%) were rejected at this stage.

However, the great strength of the AMBRE Project has been to provide a homogeneous analysis of the targetted subsample within an otherwise inhomogeneous dataset and thus the analysis of the FEROS, HARPS and UVES datasets can be considered as a combined quasi-homogeneous database. For the UVES FGK stars, the stellar sample for which the AMBRE Grid has been optimised, all four stellar parameters (effective temperature, surface gravity, metallicity and alpha element to iron ratio abundances) are reported for 10,212 ($\sim$19.7\%) of the UVES spectra corresponding to $\sim$3086 stars. For a further 438 spectra ($\sim$0.8\%) effective temperature, surface gravity and metallicity are reported (corresponding to $\sim$350 stars), while just effective temperature is reported for another 1,753 spectra ($\sim$3.4\%) corresponding to $\sim$1119 stars. Hence AMBRE:UVES has successfully extracted the FGK sample from within the UVES archive dataset by homogeneous, blind analysis.

Furthermore, the radial velocity determination was subject to less rejection criteria and so velocities with an error less than 10~kms$^{-1}$ have been reported. This resulted in radial velocities for 36,881 ($\sim$71.0\%) of the spectra (for RED L \& U spectra considered together).

Combined with the present paper, a total of $\sim$110,000 spectra have now been homogeneously parameterised by the AMBRE Project with the full parameter set of V$_{rad}$, $T_{\textrm{eff}}$, $\log g$, [M/H], [$\alpha$/Fe] and associated internal and external errors. About twice more spectra have been partly parameterised with V$_{rad}$ and $T_{\textrm{eff}}$ with associated errors.

The last stage of the AMBRE Project is the analysis of approximately the same amount of spectra again from within the GIRAFFE archived sample, which will be carried out in the near future.

\begin{acknowledgements}
The AMBRE Project team members would like to thank ESO, OCA and CNES for their financial support of this project. CCW would like to thank the IoA for their financial support. This work was partly supported by the European Union FP7 programme through ERC grant number 320360 and by the Leverhulme Trust through grant RPG-2012-541. Part of the calculations have been performed using the OCA/SIGAMM mesocentre. J.~C.~Gazzano is thanked for his help in some preliminary parts of the analysis. We would like to thank C.~Melo for use of the radial velocity programme, and also L.~Pasquini for initiating the project, as well as M.~Romaniello and J.~Melnick for their help within ESO.

This research has made use of the SIMBAD database, operated at CDS, Strasbourg, France, as well as the NASA/IPAC Infrared Science Archive, which is operated by the Jet Propulsion Laboratory, California Institute of Technology, under contract with the National Aeronautics and Space Administration.

The authors would like to thank the referee for the careful reading and comments which greatly improved this paper.
   
\end{acknowledgements}

%\bibliographystyle{aa}
%Included for Gather Purpose only:
%input "\home\cclare\AMBRE\UVES\Paper\Bibliography\matisse.bib"
%\setlinespacing{1.44}
%\bibliography{/home/cclare/AMBRE/UVES/Paper/Bibliography/matisse}

\begin{thebibliography}{23}
\expandafter\ifx\csname natexlab\endcsname\relax\def\natexlab#1{#1}\fi

\bibitem[{{Allende Prieto} {et~al.}(2004){Allende Prieto}, {Barklem},
  {Lambert}, \& {Cunha}}]{Allende-Prieto2004}
{Allende Prieto}, C., {Barklem}, P.~S., {Lambert}, D.~L., \& {Cunha}, K. 2004,
  A\&A, 420, 183

\bibitem[{{Bagnulo} {et~al.}(2003){Bagnulo}, {Jehin}, {Ledoux}, {Cabanac},
  {Melo}, {Gilmozzi}, \& {The ESO Paranal Science Operations Team}}]{UVESPOP}
{Bagnulo}, S., {Jehin}, E., {Ledoux}, C., {et~al.} 2003, The Messenger, 114, 10

\bibitem[{{Bijaoui} {et~al.}(2008){Bijaoui}, {Recio-Blanco}, \& {de
  Laverny}}]{Bijaoui2008}
{Bijaoui}, A., {Recio-Blanco}, A., \& {de Laverny}, P. 2008, in American
  Institute of Physics Conference Series, Vol. 1082, American Institute of
  Physics Conference Series, ed. {C.~A.~L.~Bailer-Jones}, 54--60

\bibitem[{{de Laverny} {et~al.}(2013){de Laverny}, {Recio-Blanco}, {Worley},
  {De Pascale}, {Hill}, \& {Bijaoui}}]{deLaverny2013}
{de Laverny}, P., {Recio-Blanco}, A., {Worley}, C.~C., {et~al.} 2013, The
  Messenger, 153, 18

\bibitem[{{de Laverny} {et~al.}(2012){de Laverny}, {Recio-Blanco}, {Worley}, \&
  {Plez}}]{delaverny2012}
{de Laverny}, P., {Recio-Blanco}, A., {Worley}, C.~C., \& {Plez}, B. 2012,
  \aap, 544, A126

\bibitem[{{De Pascale} {et~al.}(2014){De Pascale}, {Worley}, {de Laverny},
  {Recio-Blanco}, {Hill}, \& {Bijaoui}}]{DePascale2014}
{De Pascale}, M., {Worley}, C.~C., {de Laverny}, P., {et~al.} 2014, \aap, 570,
  A68

\bibitem[{{Dekker} {et~al.}(2000){Dekker}, {D'Odorico}, {Kaufer}, {Delabre}, \&
  {Kotzlowski}}]{Dekker2000}
{Dekker}, H., {D'Odorico}, S., {Kaufer}, A., {Delabre}, B., \& {Kotzlowski}, H.
  2000, in Society of Photo-Optical Instrumentation Engineers (SPIE) Conference
  Series, Vol. 4008, Society of Photo-Optical Instrumentation Engineers (SPIE)
  Conference Series, ed. {M.~Iye \& A.~F.~Moorwood}, 534--545

\bibitem[{{Drosg}(2009)}]{Drosg2009}
{Drosg}, M. 2009, Dealing with Uncertainties: A Guide to Error Analysis, 2nd
  edn. (Springer Science \& Business Media)

\bibitem[{{Gazzano} {et~al.}(2010){Gazzano}, {de Laverny}, {Deleuil},
  {Recio-Blanco}, {Bouchy}, {Moutou}, {Bijaoui}, {Ordenovic}, {Gandolfi}, \&
  {Loeillet}}]{Gazzano2010}
{Gazzano}, J.-C., {de Laverny}, P., {Deleuil}, M., {et~al.} 2010, \aap, 523,
  A91+

\bibitem[{{Gilmore} {et~al.}(2012){Gilmore}, {Randich}, {Asplund}, {Binney},
  {Bonifacio}, {Drew}, {Feltzing}, {Ferguson}, {Jeffries}, {Micela}, \&
  et~al.}]{Gilmore2012}
{Gilmore}, G., {Randich}, S., {Asplund}, M., {et~al.} 2012, The Messenger, 147,
  25

\bibitem[{{Gustafsson} {et~al.}(2008){Gustafsson}, {Edvardsson}, {Eriksson},
  {J{\o}rgensen}, {Nordlund}, \& {Plez}}]{Gustafsson2008}
{Gustafsson}, B., {Edvardsson}, B., {Eriksson}, K., {et~al.} 2008, A\&A, 486,
  951

\bibitem[{{Hinkle} {et~al.}(2000){Hinkle}, {Wallace}, {Valenti}, \&
  {Harmer}}]{ArcturusAtlasHinkle}
{Hinkle}, K., {Wallace}, L., {Valenti}, J., \& {Harmer}, D. 2000, {Visible and
  Near Infrared Atlas of the Arcturus Spectrum 3727-9300 A}

\bibitem[{{Jofr{\'e}} {et~al.}(2014){Jofr{\'e}}, {Heiter}, {Soubiran},
  {Blanco-Cuaresma}, {Worley}, {Pancino}, {Cantat-Gaudin}, {Magrini},
  {Bergemann}, {Gonz{\'a}lez Hern{\'a}ndez}, {Hill}, {Lardo}, {de Laverny},
  {Lind}, {Masseron}, {Montes}, {Mucciarelli}, {Nordlander}, {Recio Blanco},
  {Sobeck}, {Sordo}, {Sousa}, {Tabernero}, {Vallenari}, \& {Van
  Eck}}]{Jofre2014}
{Jofr{\'e}}, P., {Heiter}, U., {Soubiran}, C., {et~al.} 2014, \aap, 564, A133

\bibitem[{{Kordopatis} {et~al.}(2013){Kordopatis}, {Gilmore}, {Steinmetz},
  {Boeche}, {Seabroke}, {Siebert}, {Zwitter}, {Binney}, {de Laverny},
  {Recio-Blanco}, {Williams}, {Piffl}, {Enke}, {Roeser}, {Bijaoui}, {Wyse},
  {Freeman}, {Munari}, {Carrillo}, {Anguiano}, {Burton}, {Campbell}, {Cass},
  {Fiegert}, {Hartley}, {Parker}, {Reid}, {Ritter}, {Russell}, {Stupar},
  {Watson}, {Bienaym{\'e}}, {Bland-Hawthorn}, {Gerhard}, {Gibson}, {Grebel},
  {Helmi}, {Navarro}, {Conrad}, {Famaey}, {Faure}, {Just}, {Kos}, {Matijevi{\v
  c}}, {McMillan}, {Minchev}, {Scholz}, {Sharma}, {Siviero}, {de Boer}, \& {{\v
  Z}erjal}}]{Kordopatis2013}
{Kordopatis}, G., {Gilmore}, G., {Steinmetz}, M., {et~al.} 2013, \aj, 146, 134

\bibitem[{{Kordopatis} {et~al.}(2011){Kordopatis}, {Recio-Blanco}, {de
  Laverny}, {Bijaoui}, {Hill}, {Gilmore}, {Wyse}, \&
  {Ordenovic}}]{Kordopatis2011}
{Kordopatis}, G., {Recio-Blanco}, A., {de Laverny}, P., {et~al.} 2011, \aap,
  535, A106

\bibitem[{{Kupka} {et~al.}(1999){Kupka}, {Piskunov}, {Ryabchikova}, {Stempels},
  \& {Weiss}}]{Kupka1999}
{Kupka}, F., {Piskunov}, N., {Ryabchikova}, T.~A., {Stempels}, H.~C., \&
  {Weiss}, W.~W. 1999, \aaps, 138, 119

\bibitem[{{Recio-Blanco} {et~al.}(2006){Recio-Blanco}, {Bijaoui}, \& {de
  Laverny}}]{Recio-Blanco2006}
{Recio-Blanco}, A., {Bijaoui}, A., \& {de Laverny}, P. 2006, MNRAS, 370, 141

\bibitem[{{Recio-Blanco} {et~al.}(2016){Recio-Blanco}, {de Laverny}, {Allende
  Prieto}, {Fustes}, {Manteiga}, {Arcay}, {Bijaoui}, {Dafonte}, {Ordenovic}, \&
  {Ordo{\~n}ez Blanco}}]{Recio-Blanco2016}
{Recio-Blanco}, A., {de Laverny}, P., {Allende Prieto}, C., {et~al.} 2016,
  \aap, 585, A93

\bibitem[{{Soubiran} {et~al.}(2010){Soubiran}, {Le Campion}, {Cayrel de
  Strobel}, \& {Caillo}}]{Soubiran2010}
{Soubiran}, C., {Le Campion}, J., {Cayrel de Strobel}, G., \& {Caillo}, A.
  2010, \aap, 515, A111+

\bibitem[{{Steinmetz} {et~al.}(2006){Steinmetz}, {Zwitter}, {Siebert},
  {Watson}, {Freeman}, {Munari}, {Campbell}, {Williams}, {Seabroke}, {Wyse},
  {Parker}, {Bienaym{\'e}}, {Roeser}, {Gibson}, {Gilmore}, {Grebel}, {Helmi},
  {Navarro}, {Burton}, {Cass}, {Dawe}, {Fiegert}, {Hartley}, {Russell},
  {Saunders}, {Enke}, {Bailin}, {Binney}, {Bland-Hawthorn}, {Boeche}, {Dehnen},
  {Eisenstein}, {Evans}, {Fiorucci}, {Fulbright}, {Gerhard}, {Jauregi}, {Kelz},
  {Mijovi{\'c}}, {Minchev}, {Parmentier}, {Pe{\~n}arrubia}, {Quillen}, {Read},
  {Ruchti}, {Scholz}, {Siviero}, {Smith}, {Sordo}, {Veltz}, {Vidrih}, {von
  Berlepsch}, {Boyle}, \& {Schilbach}}]{Steinmetz2006}
{Steinmetz}, M., {Zwitter}, T., {Siebert}, A., {et~al.} 2006, \aj, 132, 1645

\bibitem[{{Wallace} {et~al.}(1998){Wallace}, {Hinkle}, \&
  {Livingston}}]{SolarAtlasHinkle}
{Wallace}, L., {Hinkle}, K., \& {Livingston}, W. 1998, {An atlas of the
  spectrum of the solar photosphere from 13,500 to 28,000 cm-1 (3570 to 7405
  A)}

\bibitem[{{Worley} {et~al.}(2012){Worley}, {de Laverny}, {Recio-Blanco},
  {Hill}, {Bijaoui}, \& {Ordenovic}}]{Worley2012}
{Worley}, C.~C., {de Laverny}, P., {Recio-Blanco}, A., {et~al.} 2012, \aap,
  542, A48

\bibitem[{{Zucker} {et~al.}(2012){Zucker}, {de Silva}, {Freeman},
  {Bland-Hawthorn}, \& {Hermes Team}}]{Zucker2012}
{Zucker}, D.~B., {de Silva}, G., {Freeman}, K., {Bland-Hawthorn}, J., \&
  {Hermes Team}. 2012, in Astronomical Society of the Pacific Conference
  Series, Vol. 458, Galactic Archaeology: Near-Field Cosmology and the
  Formation of the Milky Way, ed. W.~{Aoki}, M.~{Ishigaki}, T.~{Suda},
  T.~{Tsujimoto}, \& N.~{Arimoto}, 421

\end{thebibliography}

%\clearpage

\appendix
\section{Description of ESO Table for AMBRE:UVES}
\begin{sidewaystable*}
\tabcolsep=0.08cm
\vspace{18cm}
%\begin{landscape}
%\begin{table*}[tbp]
\caption{Description of columns in the table of UVES stellar parameters delivered to ESO}
\label{tab:esotab_descrip}
\centering
\begin{tabular}{llccl}
\hline\hline
KeyWord & Definition & Range of Values & Null value & Determination \\ 
\hline
{\scriptsize DP\_ID} & ESO data set identifier &  &  &  \\ 
{\scriptsize OBJECT} & Object designation as read in ORIGFILE &  &  &  \\ 
{\scriptsize TARG\_NAME} & Target designation as read in ORIGFILE &  &  &  \\ 
{\scriptsize RAJ2000} & Telescope pointing (right ascension, J2000) & deg &  &  \\ 
{\scriptsize DEJ2000}  & Telescope pointing (declination, J2000) & deg &  &  \\ 
{\scriptsize MJD\_OBS} & Start of observation date & Julian Day &  &  \\ 
{\scriptsize EXPTIME} & Total integration time & sec &  &  \\ 
{\scriptsize SNR} & Signal-to-Noise Ratio as estimated by the pipeline & 0-$\infty$ & NaN &  \\ 
{\scriptsize SNR\_FLAG} & Signal-to-Noise Ratio quality flag & C,R &  & C=Crude estimate from SPA$^*$, R=Refined estimate from SPC$^\#$ \\ 
{\scriptsize EXTREME\_EMISSION\_LINE\_FLAG} & Detection of extreme emission lines. & T,F &  & T=True: detection therefore no analysis carried out,  \\ 
 &  &  &  & F=False: no detection therefore analysis carried out \\ 
{\scriptsize EMISSION\_LINE\_FLAG} & Detection of some emission lines & T,F &  & T=True: some emission lines detected but analysis carried out,  \\ 
 &  &  &  & F=False: no detection therefore analysis carried out \\ 
{\scriptsize MEANFWHM\_LINES} & Mean FWHM of absorption lines & 0-0.33 & NaN & FWHM measured from spectral features (m\AA) \\ 
{\scriptsize MEANFWHM\_LINES\_FLAG} & Flag on the mean FWHM & T,F &  & T=True: FWHM $>$ 0.33 or $<$ 0.11. Default FWHM values used \\ 
 &  &  &  & F=False: FWHM $<$ 0.33, $>$ 0.11 \\ 
{\scriptsize VRAD} & Stellar radial velocity & -500 to +500 & NaN & Units=kms$^{-1}$ \\ 
{\scriptsize ERR\_VRAD} & Error on the radial velocity & 0-$\infty$ & NaN & If $\sigma_{vrad} > 10$, null value used for all stellar parameters. Units=kms$^{-1}$ \\ 
{\scriptsize VRAD\_CCF\_FWHM}  & FWHM of the CCF between the spectrum and the binary mask & 0-$\infty$ & NaN & Units=kms$^{-1}$ \\ 
{\scriptsize VRAD\_FLAG} & Quality flag on the radial velocity analysis & 0,1,2,3,4,5 & -99 & 0=Excellent determination...5=Poor determination \\ 
{\scriptsize TEFF} & Stellar effective temperature ($T_{\textrm{eff}}$)  & 3000-7625 & NaN & Units=K. Null value used if $T_{\textrm{eff}}$ is outside accepted parameter \\ 
 & as estimated by the pipeline &  &  & limits or if the spectrum is rejected due to quality flags. \\ 
{\scriptsize ERR\_INT\_TEFF} & Effective temperature internal error & 0-$\infty$ & NaN & Units=K. Square root of quadrature sum of internal errors  \\ 
 &  &  &  & ($\sigma(T_{\textrm{eff}})_{int,snr}$, $\sigma(T_{\textrm{eff}})_{int,vrad}$ \& $\sigma(T_{\textrm{eff}})_{int,norm}$ \\ 
{\scriptsize ERR\_EXT\_TEFF} & Effective temperature external error & 120 & NaN & Units=K. Maximum expected error due to external sources \\ 
{\scriptsize LOG\_G} & Stellar surface gravity (log g) as estimated by the pipeline & 1-4.9 & NaN & Units=dex. Null value used if $\log g$ is outside accepted parameter \\ 
 &  &  &  & limits or if the spectrum is rejected due to quality flags. \\ 
{\scriptsize ERR\_INT\_LOG\_G} & Surface gravity internal error & 0-$\infty$ & NaN & Units=dex. Square root of quadrature sum of internal errors  \\ 
 &  &  &  & ($\sigma(\log g)_{int,snr}$, $\sigma(\log g)_{int,vrad}$ \& $\sigma(\log g)_{int,norm}$ \\ 
{\scriptsize ERR\_EXT\_LOG\_G} & Surface gravity external error & 0.2 & NaN & Units=dex. Maximum expected error due to external sources \\ 
{\scriptsize M\_H} & Mean metallicity [M/H] as estimated by the pipeline & 0-$\infty$ & NaN & Units=dex. Null value used if [M/H] is outside accepted parameter \\ 
 &  &  &  & limits or if the spectrum is rejected due to quality flags. \\ 
{\scriptsize ERR\_INT\_M\_H} & Mean metallicity internal error & 0-$\infty$ & NaN & Units=dex. Square root of quadrature sum of internal errors  \\ 
 &  &  &  & ($\sigma(\textrm{[M/H]})_{int,snr}$, $\sigma(\textrm{[M/H]})_{int,vrad}$ \& $\sigma(\textrm{[M/H]})_{int,norm}$ \\ 
{\scriptsize ERR\_EXT\_M\_H}  & Mean metallicity external error & 0.1 & NaN & Units=dex. Maximum expected error due to external sources \\ 
{\scriptsize ALPHA} & $\alpha$-elements over iron enrichment ([$\alpha$/Fe])  & -0.4 - 0.4 & NaN & Units=dex. Null value used if [$\alpha$/Fe] is outside accepted parameter  \\ 
 & as estimated by the pipeline &  &  & limits or if the spectrum is rejected due to quality flags. \\ 
{\scriptsize ERR\_INT\_ALPHA} & $\alpha$-elements over iron enrichment internal error & 0-$\infty$ & NaN & Units=dex. Square root of quadrature sum of internal errors  \\ 
 &  &  &  & ($\sigma(\textrm{[$\alpha$/Fe]})_{int,snr}$, $\sigma(\textrm{[$\alpha$/Fe]})_{int,vrad}$ \& $\sigma(\textrm{[$\alpha$/Fe]})_{int,norm}$ \\ 
{\scriptsize ERR\_EXT\_ALPHA} & $\alpha$-elements over iron enrichment external error & 0.1 & NaN & Units=dex. Maximum expected error due to external sources \\ 
{\scriptsize CHI2} & log($\chi^2$) of the fit between the observed and the & 0-$\infty$ & NaN & Goodness of fit between final normalised  \\ 
 & reconstructed synthetic spectrum at the MATISSE parameters &  &  & and final reconstructed spectra \\ 
{\scriptsize CHI2\_FLAG} & Quality flag on the fit between the observed and the  & 0,1,2  & -99 & 0=Good fit...2=Poor fit \\ 
 & reconstructed synthetic spectrum at the MATISSE parameters &  &  &  \\ 
{\scriptsize ORIGFILE}  & ESO filename of the original spectrum being analysed &  &  &  \\ 
{\scriptsize setup} & Standard UVES setup of Obbservation &  &  &  \\ 
\hline
 &  &  &  & * = Spectral Processing B \\ 
 &  &  &  & \# = Spectral Processing C \\ 
 &  &  &  &  \\ 
\hline
\end{tabular}
%\end{table*}
%\end{landscape}
\end{sidewaystable*}

\end{document}